\documentstyle[]{article}

\def\hybrid{\topmargin 0pt      \oddsidemargin 0pt
        \headheight 0pt \headsep 0pt
        \textwidth 6.25in       % A4 paper
        \textheight 9.5in       % A4 paper
        \marginparwidth 0.0in
        \parskip 5pt plus 1pt   \jot = 1.5ex}
\catcode`\@=11
\def\marginnote#1{}

\newcount\hour
\newcount\minute
\newtoks\amorpm
\hour=\time\divide\hour by60
\minute=\time{\multiply\hour by60 \global\advance\minute by-\hour}
\edef\standardtime{{\ifnum\hour<12 \global\amorpm={am}%
        \else\global\amorpm={pm}\advance\hour by-12 \fi
        \ifnum\hour=0 \hour=12 \fi
        \number\hour:\ifnum\minute<10 0\fi\number\minute\the\amorpm}}
\edef\militarytime{\number\hour:\ifnum\minute<10 0\fi\number\minute}

\def\draftlabel#1{{\@bsphack\if@filesw {\let\thepage\relax
   \xdef\@gtempa{\write\@auxout{\string
      \newlabel{#1}{{\@currentlabel}{\thepage}}}}}\@gtempa
   \if@nobreak \ifvmode\nobreak\fi\fi\fi\@esphack}
        \gdef\@eqnlabel{#1}}
\def\@eqnlabel{}
\def\@vacuum{}
\def\draftmarginnote#1{\marginpar{\raggedright\scriptsize\tt#1}}

\def\draftlabel#1{{\@bsphack\if@filesw {\let\thepage\relax
   \xdef\@gtempa{\write\@auxout{\string
      \newlabel{#1}{{\@currentlabel}{\thepage}}}}}\@gtempa
   \if@nobreak \ifvmode\nobreak\fi\fi\fi\@esphack}
        \gdef\@eqnlabel{#1}}
\def\@eqnlabel{}
\def\@vacuum{}
\def\draftmarginnote#1{\marginpar{\raggedright\scriptsize\tt#1}}

\def\draft{\oddsidemargin -.5truein
        \def\@oddfoot{\sl preliminary draft \hfil
        \rm\thepage\hfil\sl\today\quad\militarytime}
        \let\@evenfoot\@oddfoot \overfullrule 3pt
        \let\label=\draftlabel
        \let\marginnote=\draftmarginnote
   \def\@eqnnum{(\theequation)\rlap{\kern\marginparsep\tt\@eqnlabel}%
\global\let\@eqnlabel\@vacuum}  }

\def\numberbysection{\@addtoreset{equation}{section}
        \def\theequation{\thesection.\arabic{equation}}}

\def\underline#1{\relax\ifmmode\@@underline#1\else
        $\@@underline{\hbox{#1}}$\relax\fi}

\def\titlepage{\@restonecolfalse\if@twocolumn\@restonecoltrue\onecolumn
     \else \newpage \fi \thispagestyle{empty}\c@page\z@
        \def\thefootnote{\fnsymbol{footnote}} }

\def\endtitlepage{\if@restonecol\twocolumn \else  \fi
        \def\thefootnote{\arabic{footnote}}
        \setcounter{footnote}{0}}  %\c@footnote\z@ }
\relax

\numberbysection
\hybrid

\def\beq{\begin{equation}}
\def\eeq{\end{equation}}
\def\p{\partial}
\def\G{\Gamma}

\newtheorem{th}{Theorem}[section]

\newtheorem{cor}{Corollary}[section]
\newtheorem{lem}{Lemma}[section]

\def\square{\hfill
{\vrule height6pt width6pt depth1pt} \break \vspace{.01cm}}

\begin{document}

\begin{titlepage}

\title{Spin generalization of the Ruijsenaars-Schneider model,
non-abelian 2D Toda chain and representations of Sklyanin algebra}

\author{I. Krichever \thanks{ Landau Institute for Theoretical Physics
 Kosygina str. 2, 117940 Moscow, Russia}
\and A. Zabrodin
\thanks{Joint Institute of Chemical Physics, Kosygina str. 4, 117334,
Moscow, Russia and ITEP, 117259, Moscow, Russia}}
\date{May 1995}
\maketitle
\begin{abstract}
Action-angle type variables for spin generalizations of the elliptic
Ruijsenaars-Schneider system are constructed. The equations of motion
of these systems are solved in terms of Riemann theta-functions.
It is proved that these systems are isomorphic to special elliptic
solutions of the non-abelian 2D Toda chain. A connection between the
finite gap solutions of solitonic equations and representations of
the Sklyanin algebra is revealed and discrete analogs of the
Lame operators are introduced. A simple way to construct representations
of the Sklyanin algebra by difference operators is suggested.
\end{abstract}

\vfill

\end{titlepage}
\newpage

\section{Introduction}
In a sense this paper is a preliminary result of our recent attempt to
analyse representations of the Sklyanin algebra. This is the algebra
with four generators $S_0, S_{\alpha},\ \alpha=1,2,3$, subject to
homogeneous quadratic relations \beq [S_0,S_{\alpha}]_-=iJ_{\beta
\gamma} [S_{\beta},S_{\gamma}]_+, \label{1} \eeq \beq
[S_{\alpha},S_{\beta}]_-=i[S_0,S_{\gamma}]_+ \label{2}
\eeq
($[A,B]_{\pm}=AB\pm BA$, a triple of Greek indices $\alpha, \beta,
\gamma$ in (\ref{1}, \ref{2}) stands for any {\it cyclic}
permutation of $(1,2,3)$).
Structure constants of the algebra $J_{\alpha \beta}$ have the form
\beq
J_{\alpha \beta}={J_{\beta}-J_{\alpha}\over J_{\gamma}} ,\label{3}
\eeq
where $J_{\alpha}$ are arbitrary constants. Therefore, the relations
(\ref{1}-\ref{3}) define a two-paramet\-ric family of quadratic algebras.
These relations (\ref{1}-\ref{3}) were introduced in the
paper \cite{skl1} as the minimal set of conditions under which
operators

\beq
L(u)=\sum_{a=0}^3 W_{a}(u)S_{a}\otimes \sigma_{a}, \label{4}
\eeq
satisfy the equation
\beq
R^{23}(u-v) L^{13}(u) L^{12}(v) = L^{12}(v) L^{13} (u) R^{23}(u-v). \label{5}
\eeq
Here
$\sigma_{\alpha}$
are Pauli matrices,
$\sigma_0$
is the unit matrix;
$W_{a}(u)=W_{a}(u|\eta,\tau),\ a=0,\ldots,3$
are functions of the variable $u$ with parameters $\eta$ and $\tau$:
\beq
W_a(u)={\theta_{a+1}(u)\over \theta_{a+1}(\eta /2)} \label{9}
\eeq
($\theta_a(x)=\theta _{a}(x|\tau)$ are standard Jacobi theta-functions
with characteristics and the modular parameter $\tau$; for the
definitions see Appendix to Section 6);
\beq
R(u)=\sum_{a=0}^3 W_{a}(u+\frac{\eta}{2})\sigma_{a}\otimes
\sigma_{a} \label{6}
\eeq
is the elliptic solution of the quantum Yang-Baxter equation
\beq
R^{23}(u-v) R^{13}(u) R^{12}(v) =R^{12}(v) R^{13} (u) R^{23}(u-v) \label{7}
\eeq
that corresponds to the 8-vertex model. (The limit $\tau \to 0$
yields the $R$-matrix
of the 6-vertex model.)

In (\ref{5}), (\ref{7}) we use the following standard notation.
For any module $M$ over the Sklyanin algebra
eq. (\ref{4}) defines an operator in the tensor product
$M\otimes {\bf C}^2$. The operator $L^{13}(u) \ (L^{12}(u))$
in the tensor product $M\otimes {\bf C}^2\otimes {\bf C}^2$
acts as $L(u)$ on the first and the third spaces and as the identity
operator on the second one (acts as $L(u)$
on the first and the second spaces and as the identity operator on
the third one).
Similarly, $R^{23}$ acts identically on $M$ and coincides with
the operator (\ref{6}) on the last two spaces.

Classification of discrete quantum systems solvable by the quantum
inverse scattering method (see the reviews \cite{ft},
\cite{kulskl}, \cite{izkor}) is equivalent to
solving eq. (\ref{5}),
where $R(u)$ is a fixed solution to the Yang-Baxter
equation (\ref{7}).

More general elliptic solutions to (\ref{7}) were found
in \cite{bel}. Corresponding generalizations of the Sklyanin algebra
were introduced in
\cite{feig}, \cite{artin}. At present time only the simplest
finite-dimensional representations of these generalized Sklyanin algebras
are known. It would be very interesting to construct
representations of these algebras in terms of difference operators
similar to those found in \cite{skl2} for the original Sklyanin algebra.

As it was shown in the paper \cite{skl2}, the operators
$S_a,\ a=0,\ldots,3$ admit representations in the form of
second order difference operators acting in the space
of meromorphic functions $f(x)$ of one complex variable $x$.
One of the series of such representations has the form
\beq
(S_a f)(x)=(i)^{\delta_{2,a}}\theta _{a+1}(\eta /2)
{\theta _{a+1}(x-\ell \eta) f(x+\eta)-
\theta _{a+1}(-x-\ell \eta)f(x-\eta)\over \theta_{1}(x)}.
\label{10}
\eeq

By a straightforward but tedious computation one can check that
for any  $\tau,\ \eta, \ \ell$ the operators
(\ref{10}) satisfy commutation relations
(\ref{1}-\ref{3}), the values of the structure constants being
\beq
J_{\alpha}={\theta_{\alpha +1}(\eta)\theta_{\alpha +1}(0)\over
\theta_{\alpha +1}^2(\eta/2)}. \label{13}
\eeq
Therefore, the values of $\tau$ and $\eta$ parametrize the structure
constants, while $\ell$ is the parameter of the representation.
Note that the original Sklyanin's parameter denoted by $\eta$ in the
paper \cite{skl2} is equal to {\it half} of our
$\eta $ entering (\ref{9}),
(\ref{6}), (\ref{10}) and (\ref{13}).

Putting $f_n=f(n\eta+x_0)$, to the operators (\ref{10}) we assign
difference Schr\"odinger operators
\beq
S_a f_n=A_n^a f_{n+1}+B_n^a f_{n-1}  \label{14}
\eeq
with quasiperiodic coefficients. The spectrum of a generic
operator of this form in the space
$l^2({\bf Z})$ (square integrable sequences $f_n$) has the structure
of the Cantor set type.
If $\eta$ is a rational number,
$\eta=p/q$, then operators (\ref{14}) have $q$-periodic coefficients.
In general $q$-periodic difference Schr\"odinger operators have $q$
unstable bands in the spectrum.

In Sect. 5 of this paper we show that the spectral
properties of the operator $S_0$
given by eq. (\ref{10}) are in this sense extremely unusual:
\begin{th}
The operator $S_0$ given by eq. (\ref{10}) for positive integer values of
"spin" $\ell$
and arbitrary $\eta$ has $2\ell$ unstable bands in the spectrum.
Its Bloch functions are parametrized by points of a hyperelliptic
curve of genus $2\ell$ defined by the equation
\beq
y^2=R(\varepsilon)=\prod_{i=1}^{2\ell +1}(\varepsilon^2-\varepsilon_i^2).
\label{15} \eeq
Bloch eighenfunctions $\psi(x, \pm \varepsilon_i)$ of the operator $S_0$ at
the edges of bands span an invariant functional subspace for all
operators $S_a$.  The corresponding $4\ell +2$-dimensional representation
of the Sklyanin algebra is a direct sum of two equivalent
$2\ell +1$-dimensional representations of the Sklyanin algebra.  \end{th}
{\bf Remark.} In section 5 we show  that there is a unique
choice of signs for $\varepsilon_i$ such that the space of
{\it irreducible} representation of the Sklyanin algebra is spanned by
Bloch eigenfunctions $\psi(x,\varepsilon_i)$. Unfortunately,
at this stage we do not
know at this stage any explicit constructive description of the
corresponding splitting of the edges in two parts. We
conjecture that for real structure constants (when all $\varepsilon_i$
are real) one has to choose all positive edges of bands:
$\varepsilon_i>0$.

This theorem indicates a connection between representations of the
Sklyanin algebra and the theory of finite gap integration of solitonic
equations. (The theory of finite difference Schr\"odinger operators
\cite{dmn}, \cite{kr1}, \cite{kr2}, \cite{dattan} was developed
in the context of solving the Toda chain and difference
KdV equations.) Furthermore, this theorem suggests that $S_0$
is the proper difference analog of the classical Lame operator
\beq L=-{d^2\over
dx^2}+\ell (\ell +1)\wp(x) ,\label{16}
\eeq
which can be obtained from $S_0$ in the limit
$\eta\to 0$. Finite gap properties of higher Lame operators (for arbitrary
integer values of $\ell$) were established in \cite{ince}.

In Sect. 6 of this paper we suggest a relatively simple way to derive
the realization (\ref{10}) of the Sklyanin algebra by difference
operators. This approach partially explains the origin of these
operators. The basic tool is a key property of the elementary $R$-matrix
(\ref{6}), which was used by Baxter in his solution of the
eight-vertex model and called by him "pair-propagation through
a vertex" \cite{Baxter}. A suitable generalization of this
property for the arbitrary spin $L$-operator (\ref{4}) leads to
formulas (\ref{10}). This derivation needs much less amount of
computations than the direct substitution of the operators (\ref{10})
into the commutation relations (\ref{1}, \ref{2}). This method
gives automatically the three representation series obtained by
Sklyanin and an extra one which, presumably, was unknown.

In the paper \cite{mma} a remarkable connection between the motion of
poles of the elliptic solutions of KdV equation (which are isospectral
deformations of the higher Lame potentials) and the Calogero-Moser
dynamical system was revealed. As it has been shown in
\cite{kr3}, \cite{chud}, this relation becomes an isomorphism
in the case of the elliptic solutions of the Kadomtsev-Petviashvili (KP)
equation. The methods of finite gap integration of the KP
equation were applyed to the Calogero-Moser system in the paper
\cite{kr4}, where the complete solution in
terms of Riemann theta-functions was
obtained. These results have been extended to spin generalizations of
the Calogero-Moser system in the paper
\cite{bab}.

The main goal of this work is to extend this theory to elliptic solutions
of the two-dimensional (2D) Toda chain and its non-abelian analogs. The
equations of the 2D Toda chain have the form
\beq
\p_+\p_- \varphi_n=e^{\varphi_n-\varphi_{n-1}}-e^{\varphi_{n+1}-\varphi_n},
\ \ \p_{\pm}={\p \over \p t_{\pm}}.
\label{17}
\eeq
Let us consider {\it elliptic solutions with respect to the discrete
variable} $n$, specifically,
\beq
\varphi_n(t_+,t_-)=\varphi(n\eta+x_0,t_+,t_-) \label{18}
\eeq
such that
\beq
c(x,t_+,t_-)=\exp (\varphi(x,t_+,t_-)-\varphi(x-\eta,t_+,t_-))\label{19}
\eeq
is an elliptic function of the variable $x$. In what follows we show that
the function\\
$(\exp \varphi)$ has the form
\beq
\exp \varphi(x,t_+,t_-)=\prod_{i=1}^n
{\sigma(x-x_i+\eta)\over \sigma(x-x_i) },\qquad x_i=x_i(t_+,t_-), \label{20}
\eeq
($\sigma(x|\omega_1 ,\omega_2)$ is the standard Weierstrass
$\sigma$-function) and the dynamics of its poles $x_i$
with respect to the time flows
$t_+, t_-$ coincides with the equations of motion of the
Ruijsenaars-Schneider dynamical system:
\cite{ruij}:  \beq \ddot x_i = \sum_{s\neq i} \dot
x_i \dot x_s (V(x_i-x_s)-V(x_s-x_i)), \label{21} \eeq
where \beq
V(x)=\zeta(x)-\zeta(x+\eta), \ \ \zeta(x)={\sigma(x)'\over \sigma(x)}.
\label{22} \eeq
This system is the relativistic analog of the Calogero-Moser model.
Hamiltonians generating the commuting $t_{\pm}$-flows have the form
\beq H_{\pm}
=\sum _{j=1}^{n}e^{\pm p_{j}}\prod _{s \ne j}^{n}\left ( \frac {\sigma
(x_j -x_s +\eta )\sigma (x_j -x_s -\eta )} {\sigma ^{2} (x_j -x_s
)}\right ) ^{1/2} \label{21a} \eeq
with canonical Poisson brackets: $\{ p_i , x_k \}= \delta _{ik}$.

Our proof of this statement allows us to construct the action-angle
variables for the system (\ref{21}) and to solve it explicitly in terms
of theta-functions.
Being applyed to the non-abelian analog of the 2D Toda chain, this approach
leads to spin generalization of the Ruijsenaars-Schneider model.

This generalized model is a system of $N$ particles on the line
with coordinates $x_i$ and internal degrees of freedom given by
$l$-dimensional vectors
$a_i=(a_{i,\alpha})$ and $l$-dimensional covectors
$b_i^+=(b_i^{\alpha})$ , $\alpha=1,\ldots,l$.
The equations of motion have the form
\beq
\ddot{x}_i = \sum_{j\neq i} (b_i^+a_j)(b_j^+a_i)(V(x_i-x_j)-V(x_j-x_i)),
\label{23}
\eeq
\beq
\dot{a}_i= \sum_{j\neq i}a_j(b_j^+a_i)V(x_i-x_j),\label{24}
\eeq
\beq
\dot{b}_i^+=-\sum_{j\neq i}b_j^+(b_i^+a_j)V(x_j-x_i). \label{25}
\eeq
The potential $V(x)$ is given by the function (\ref{22}) or its
trigonometric or rational degenerations
($V(x)=(\coth x)^{-1}-(\coth(x+\eta))^{-1}, V(x)=x^{-1}-(x-\eta)^{-1}$,
respectively). Hamiltonian formalism for this system needs special
consideration and will not be discussed in this paper.

Let us count the number of non-trivial degrees of freedom.
The original system has $2N+2Nl$ dynamical variables
$x_i$, $\dot x_i$, $a_{i,\alpha}$,
$b_i^\alpha$. The equations of motion (\ref{23}-\ref{25}) are symmetric
under rescaling
\beq
a_i\to \lambda_i a_i, \quad b_i\to {1\over\lambda_i}b_i
\label{26} \eeq
The corresponding integrals of motion have the form
$I_i=\dot x_i-(b_i^+a_i)$. Let us put them equal to zero:
\beq
\dot{x}_i=(b_i^+a_i) , \label{27}
\eeq
The reduced system is defined by $N$ extra constraints $\sum_\alpha
b_i^\alpha =1$ (they destroy the symmetry (\ref{26})). Therefore,
the phase space of the reduced system has dimension $2Nl$.
Moreover, the system has a further symmetry:
\beq a_i\to W^{-1} a_i,\quad b_i^+ \to b_i^+ W,
\label{28} \eeq
where $W$ is any matrix in $GL(r,{\bf R})$ (independent of $i$) preserving
the above condition on $b_i$'s. This means that $W$ must leave the vector
$v=(1,\cdots,1)$ invariant. Taking this symmetry into account, it is easy
to see that dimension of the completely reduced phase space ${\cal M}$ is
\beq
{\rm dim~} {\cal M} = 2\left[ Nl -{l(l-1) \over 2} \right].\label{29}
\eeq

In the next three sections of this paper we derive explicit formulas
for general solutions of the system (\ref{23}-\ref{25}) in terms of
theta-functions. We would like to stress that these formulas are
identical to those obtained
for spin generalizations of the Calogero-Moser
model in the paper
\cite{bab}. At the same time the class of auxiliary spectral curves
(in terms of which the theta-functions are constructed) is different.
These curves can be described purely in terms of algebraic geometry.

To each smooth algebraic curve $\G$ of genus $N$ it corresponds a
$N$-dimensional complex torus $J(\G)$ (Jacobian of
the curve). A pair of points $P^{\pm} \in \G$ defines a vector $U$ in
the Jacobian.  Let us consider a class of curves
having the following property: there exists
a pair of points on the curve such that the complex linear subspace
generated by the corresponding vector $U$ {\it is compact}, i.e., it is
an elliptic curve ${\cal E}_0$. This means that there exist two
complex numbers
$2\omega_{\alpha}, \ {\rm Im} \ \omega_2/\omega_1>0$, such that
$2\omega_{\alpha}U$ belongs to the lattice of periods of holomorphic
differentials on $\G$. From pure algebraic-geometrical point of view
the problem of the description of such  curves is transcendental. It
turns out that this problem has an explicit solution and algebraic
equations that define such curves can be written as a characteristic
equations for the Lax operator corresponding to the Ruijsenaars-Shneider
system.  Moreovere, it turns out that in general position  ${\cal E}_0$
intersects theta-divisor at $N$ points $x_i$ and if we move
${\cal E}_0$ in the direction that is defined by the vector $V^+$
($V^-$) tangent to $\G\in J(\G)$ at the point $P^+$ ($P^-$), then
the intersections of ${\cal E}_0$ with the theta-divisor move
according to the Ruijsenaars-Shneider dynamics. An analogous
description of spin generalisations of this system is very similar.
The corresponding curves have two sets of points $P_i^{\pm}, \
i=1,\ldots,l$ such that in the linear subspace spanned by the vectors
corresponding to each pair there exist a vector $U$ with the same
property as above.

The following remark is in order. The geometric interpretation of
integrable many body systems of Calogero-Moser-Sutherland type
consists in the representation of the models as reductions of geodesic
flows on symmetric spaces \cite{OP}. Equivalently, these models can be
obtained from free dynamics in a larger phase space possessing a rich
symmetry by means of the hamiltonian reduction \cite{KKS}. A
generalization to infinite-dimensional phase spaces (cotangent
bundles to current algebras and groups) was suggested in \cite{GN1},
\cite{GN2}. The infinite-dimensional gauge symmetry allows one to
make a reduction to finite degrees of freedom. Among systems
having appeared this way, there are Ruijsenaars-Schneider-type models and
the elliptic Calogero-Moser model.

A further generalization of this approach should consist in considering
dynamical systems on cotangent bundles to moduli spaces of stable
holomorphic vector bundles on Riemann surfaces. Such systems were
introduced by Hitchin in the paper \cite{Hi}, where their integrability
was proved. An attempt to identify the known many body integrable
systems in terms of the abstract formalism developed by Hitchin
was recently made in \cite{Nekras}. To do this, it is necessary to
consider vector bundles on algebraic curves with singular points.
It turns out that the class of integrable systems corersponding to
the Riemann sphere with marked points includes spin generalizations
of the Calogero-Moser model as well as integrable Gaudin magnets
\cite{Gaudin} (see also \cite{ER}).

However, Hitchin's approach, explaining the algebraic-geometrical
origin of integrable
systems, does not allow one to obtain explicit formulas for
solutions of equations of motion. Furthermore, in general case
any explicit form of the equations of
motion is unknown. We hope that the method suggested for the first time in
\cite{kr4}) (for the elliptic Calogero-Moser system) and further
developed in the present paper, could give an alternative approach
to Hitchin's systems; may be less invariant but yielding more
explicit formulas. We suspect that this method has not yet been
used in its full strength. Conjecturelly, to each Hitchin's system
one can assign a linear problem having solutions of a special
form (called double-Bloch solutions in this paper).
In terms of these solutions one might construct explicit
formulas solving the initial system.

Concluding the introduction, we remark that this paper can be devided
into three parts which are relatively independent. The structure of the
first one (Sects.~2~--~4) is very similar to that of the paper
\cite{bab}. Furthermore, in order to make our paper self-contained
and to stress the universal character of the method
suggested in
\cite{kr4}, we sometimes use the literal citation of the paper \cite{bab}.
At the same time we skip some technical detailes common for both cases,
trying to stress the specifics of difference equations. In the
second part (Sect. 5)
discrete analogs of Lame operators are introduced and studied. Finally,
in the third part (Sect. 6)
we give a simple derivation of difference operators
representing the Sklyanin algebra and explain the origin of these
operators. Actually, we expect a deeper connection between the three
main topics of this paper; a discussion on this point is given
in Sect. 7.

\section{Generating linear problem}

The equations of the non-abelian 2D Toda chain have the form
\beq
\p_+((\p_-g_n) g_n^{-1})=g_ng_{n-1}^{-1} -  g_{n+1}g_n^{-1}. \label{1.1}
\eeq
These equations are equivalent to the compatibility condition for the
overdetermined system of the linear problems
\beq
\p_+\psi_n(t_+,t_-)=\psi_{n+1}(t_+,t_-)+v_n(t_+,t_-) \psi_n(t_+,t_-) ,
\label{1.2}
\eeq
\beq
\p_-\psi_n(t_+,t_-)=c_n(t_+,t_-) \psi_{n-1} (t_+,t_-), \label{1.3}
\eeq
where
\beq
c_n=g_ng_{n-1}^{-1}, \qquad v_n=(\p_+g_n) g_n^{-1} \label{1.4}
\eeq
($g_n$ is $l \times l$ matrix).
Similarly to the case of the Calogero-Moser model and its spin
generalizations \cite{kr4}, \cite{bab}, the isomorphism between
the system (\ref{23}-\ref{25}) and the pole dynamics of elliptic solutions
to the non-abelian 2D Toda chain is based on the fact that the
auxiliary linear problem with elliptic coefficients has
infinite number of {\it double-Bloch} solutions.

We call a meromorphic vector-function $f(x)$ that
satisfies the following monodromy properties:
\beq
f(x+2\omega_{\alpha})=B_{\alpha} f(x), \ \  \alpha=1,2,\label{g1}
\eeq
a {\it double-Bloch function}. Here $\omega _{\alpha}$ are two
periods of an elliptic curve. The
complex numbers $B_{\alpha}$ are  called {\it Bloch
multipliers}.  (In other words, $f$ is a meromorphic section of a vector
bundle over the elliptic curve.) Any double-Bloch function can be
represented as a linear combination of elementary ones.

Let us define a function $\Phi(x,z)$ by the formula
\beq
\Phi(x,z)={\sigma(z+x+\eta)\over \sigma(z+\eta) \sigma(x)} \left[
{\sigma(z-\eta)\over \sigma(z+\eta)}\right]^{x/2\eta}. \label{1.5}
\eeq
Using addition theorems for the Weierstrass
$\sigma$-function, it is easy to check that this function satisfies
the difference analog of the Lame equation:
\beq
\Phi(x+\eta,z)+c(x)\Phi(x-\eta,z)=E(z)\Phi(x,z) , \label{1.5a} \eeq
where
\beq
c(x)={\sigma(x-\eta)\sigma(x+2\eta)\over \sigma(x+\eta)\sigma(x)}. \label{1.5b}
\eeq
Here $z$ plays the role of spectral parameter and parametrizes
eigenvalues $E(z)$ of the difference Lame operator:
\beq
E(z)={\sigma(2\eta)\over \sigma(\eta)} {\sigma(z)\over
(\sigma(z-\eta)\sigma(z+\eta))^{1/2}}. \label{1.5c}
\eeq
The Riemann surface $\hat {\G}_0$ of the function $E(z)$
is a two-fold covering of the initial elliptic curve
$\G_0$ with periods $2\omega_{\alpha}, \
\alpha=1,2$. Its genus is equal to $2$.

Considered as a function of $z$, $\Phi(x,z)$ is double-periodic:
\beq
\Phi(x,z+2\omega_{\alpha})=\Phi(x,z). \label{1.10}
\eeq
For values of $x$ such that $x/2\eta$ is an integer,
$\Phi$ is well defined meromorphic function on
$\G_0$. If $x/2\eta$ is a half-integer number, then $\Phi$
becomes single-valued on $\hat {\G}_0$.
For general values of $x$ one can define a single-valued branch of
$\Phi(x,z)$ by cutting the elliptic curve $\G_0$ between the points
$z=\pm \eta$.

As a function of $x$ the function $\Phi(x,z)$ is double-Bloch function, i.e.
\beq
\Phi(x+2\omega_{\alpha}, z)=T_{\alpha}(z) \Phi (x, z), \label{1.11}
\eeq
where Bloch multipliers are equal to
\beq
T_{\alpha}(z)=\exp (2\zeta (\omega _{\alpha})(z+\eta ))
\left (\frac{\sigma (z-\eta )}{\sigma
(z+\eta )}\right )^{\omega_{\alpha}/\eta}.
\label{1.11a}
\eeq
In the fundamental domain of the lattice defined by
$2\omega_{\alpha}$ the function $\Phi(x,z)$ has a unique
pole at the point $x=0$:
\beq
\Phi(x,z)={1\over x}+A+ O(x), \ \ A=\zeta (z+\eta )+\frac{1}{2\eta }
\ln \frac{\sigma (z-\eta )}{\sigma(z+\eta )}. \label{1.12}
\eeq
That implies that any double-Bloch function $f(x)$ with
simple poles at points $x_i$ in the fundamental domain and with
Bloch multipliers $B_{\alpha}$ such that
at least one of them is not equal to $1$
may be represented in the form:
\beq
f(x)=\sum_{i=1}^N s_i\Phi(x-x_i,z) k^{x/\eta},\label{g2}
\eeq
where $z$ and a complex number $k$ are related by
\beq
B_{\alpha}=T_{\alpha}(z) k^{2\omega_{\alpha}/\eta}. \label{g3}
\eeq
(Any pair of Bloch multipliers may be represented in the form (\ref{g3})
with an appropriate choice of parameters $z$ and $k$.)

Indeed, let $x_i, \ i=1,\ldots,m,$
be poles of $f(x)$  in the fundamental
domain of the lattice with periods
$2\omega _1 , 2\omega _2$.
Then there exist vectors $s_i$ such that the function
$$ F(x)=f(x)-\sum_{i=1}^m s_i\Phi(x-x_i,z)k^{x/\eta} $$
is {\it holomorphic} in $x$ in the fundamental domain.
It is a double-Bloch function with the same Bloch multipliers as
the function $f$. Any non-trivial double-Bloch function with at least
one of the Bloch multipliers that is not equal to $1$
has at least one pole in the fundamental domain.
Hence $F=0$.

The gauge transformation
\beq
f(x)\longmapsto \tilde f(x)=f(x)e^{ax}, \label{gn}
\eeq
where $a$ is an arbitrary constant does not change poles of any
functions and transforms a double-Bloch function into
another double-Bloch function. If $B_{\alpha}$ are Bloch multipliers for
$f$, then the Bloch multipliers for $\tilde f$ are equal to
\beq \tilde
B_1=B_1e^{2a\omega_1},\ \ \tilde B_2=B_2 e^{2a\omega_2}. \label{gn2}
\eeq
Two pairs of Bloch multipliers are said to be {\it equivalent} if they
are connected by the relation (\ref{gn2}) with
some $a$.  Note that for all equivalent pairs of Bloch
multipliers the product \beq B_{1}^{\omega_2} B_{2}^{-\omega_1}=B.
\label{gn3} \eeq is a constant depending on the equivalence class only.

\begin{th} The equations
\beq
\p_t\Psi(x,t)=\Psi(x+\eta,t) + \sum_{i=1}^N a_i (t) b_i^+(t)V(x-x_i(t))
\Psi(x,t), \label{1.6}
\eeq
\beq
-\p_t\Psi^+ (x,t)=\Psi^+(x-\eta,t)+ \Psi^+(x,t)\sum_{i=1}^N a_i (t)
b_i^+(t)V(x-x_i(t))\label{1.6a}
\eeq
have $N$ pairs of linearly independent double-Bloch solutions
$\Psi_{(s)}(x,t),\, \Psi_{(s)}^+(x,t)$ with simple poles at points
$x_i(t)$ and $(x_i(t)-\eta)$, respectively,
\beq
\Psi_{(s)}(x+2\omega_{\alpha},t)=B_{\alpha,s}\Psi_{(s)}(x,t),\ \
\Psi_{(s)}^+(x-2\omega_{\alpha},t)=B_{\alpha,s}\Psi_{(s)}(x,t), \label{g4}
\eeq
with equivalent Bloch multipliers (i.e. such that the value
\beq
B_{1,s}^{\omega_2} B_{2,s}^{-\omega_1}=B, \label{g5}
\eeq
does not depend on s)
if and only if $x_i(t)$ satisfy equations (\ref{23}) and the
vectors $a_i,\ b_i^+$ satisfy the constraints (\ref{27}) and
the system of equations
\beq
\dot{a}_i= \sum_{j\neq i}a_j(b_j^+a_i)V(x_i-x_j)-\lambda_i a_i,\label{1.8}
\eeq
\beq
\dot{b}_i^+=-\sum_{j\neq i}b_j^+(b_i^+a_j)V(x_j-x_i)+ \lambda_i
b_i^+\label{1.9}, \eeq
where $\lambda_i=\lambda_i(t)$ are scalar functions.
\end{th}
{\bf Remark.} The system (\ref{23}), (\ref{1.8}),
(\ref{1.9}) is "gauge equivalent"
to the system (\ref{23})-(\ref{25}). This means that if
$(x_i, a_i,  b_i^+)$ satisfy the equations
(\ref{23}), (\ref{1.8}), (\ref{1.9}), then $x_i$ and the vector-functions
\beq \hat a_i=a_iq_i, \ \hat b_i^+=b_iq_i^{-1},\ \
q_i=\exp(\int^t \lambda_i(t')dt') \label{1.10a}
\eeq
are solutions of the system (\ref{23}-\ref{25}).

\begin{th} If the equations (\ref{1.6}), (\ref{1.6a}) have $N$ linearly
independent double-Bloch solutions with Bloch multipliers
satisfying (\ref{g5}) then they have infinite number of them. All these
solutions
can be represented in the form:
\beq
\Psi=\sum_{i=1}^N s_i(t,k,z)\Phi(x-x_i(t),z) k^{x/\eta},\label{1.7}
\eeq
\beq
\Psi^+=\sum_{i=1}^N s_i^+(t,k,z)
\Phi(-x+x_i(t)-\eta,z)k^{-x/\eta}, \label{1.7a}
\eeq
where $s_i$ is $l$-dimensional vector
$s_i=(s_{i,\alpha})$, $s_i^+$ is $l$-dimensional covector
$s_i^+=(s_i^{\alpha})$.
The set of corresponding pairs $(z,k)$ are parametrized by points
of algebraic curve defined by the equation of the form
$$R(k,z)=k^N+\sum_{i=1}^N r_i(z) k^{N-i}=0$$
\end{th}
{\it Proof of Theorem 2.1.}
As it was mentioned above, $\Psi_{(s)}(x,t)$ (as any double-Bloch function)
may be written in the form (\ref{1.7}) with some values
of the parameters $z_s, k_s$.
The condition (\ref{g5}) implies that these parameters can
be choosen as follows:
$$z_s=z,\ \ \ s=1,\ldots,N.$$

Let us substitute the function $\Psi(x,t,z,k)$ of the form (\ref{1.7})
with this particular value of $z$ into eq. (\ref{1.6}).
Since any function with such monodromy properties has at least one pole,
it follows that the equation (\ref{1.6}) is satisfied if and
only if the right and left hand sides of this equality have the same
singular parts at the points $x=x_i$ and $x=x_i-\eta$.

Comparing the coefficients in front of $(x-x_i)^{-2}$ in
(\ref{1.6}) gives the equalities
\beq \dot{x}_i s_i= a_i(b_{i}^{+}s_i),
\label{1.13} \eeq
hence the vector $s_i$ is proportional to $a_i$:
\beq
s_{i,\alpha}(t,k,z)=c_i(t,k,z)a_{i,\alpha}(t),
\label{a1}
\eeq
and $a_i$, $b_i$ satisfy the constraints (\ref{27}). Cancellation
of the coefficients in front of
$(x-x_i +\eta)^{-1}$ gives the conditions
\beq
-ks_i+\sum_{j\neq i}a_ib_i^+ s_j\Phi(x_i-x_j-\eta ,z)=0.
\label{a2}
\eeq
Taking into account (\ref{27}) and (\ref{a1}), we rewrite them as a
matrix equation for the vector $C=(c_i)$:
\beq
(L(t,z)-kI)C=0\,,
\label{a3}
\eeq
where $I$ is the unit matrix and the Lax matrix $L(t,z)$ is defined
as follows:
\beq
L_{ij}(t,z)=(b_{i}^{+}a_j)\Phi(x_i-x_j-\eta,z).  \label{LaxL}
\eeq
Finally, cancellation of the poles
$(x-x_i)^{-1}$ gives the conditions
\beq
\dot{s}_i -\left(\sum_{j\neq
i}a_jb_j^+ V(x_i-x_j)+(A-\zeta (\eta))a_i b_{i}^{+} \right)s_i-
a_i\sum_{j\neq i}(b_i^+s_j)\Phi (x_i-x_j,z)=0.
\label{a4}
\eeq
Using
(\ref{a1}), we get the equations of motion (\ref{1.8}), where
\beq
\lambda_i (t)={\dot{c}_i\over c_i}+
(\zeta(\eta)-A)\dot{x}_i -\sum_{j\neq i}(b_i^+a_j) \Phi
(x_i-x_j,z){c_j\over c_i}.
\label{a5}
\eeq
Rewriting (\ref{a5}) in the matrix form,
\beq
(\partial_t+M(t,z))C=0\,,
\label{a6}
\eeq
we find the second operator of the Lax pair:
\beq
M_{ij}(t,z)=(-\lambda_i +(\zeta
(\eta)-A)\dot{x}_i )\delta_{ij}-
(1-\delta_{ij})b_{i}^{+}a_j \Phi (x_i -x_j ,z). \label{LaxM}
\eeq
Substituting in the same way the
vector $\Psi ^{+}$ (\ref{1.7a}) into the equation
(\ref{1.6a}), we get:
\beq
s_i^{\alpha}(t,k,z)=c_i^+ (t,k,z) b_i^{\alpha}(t),
\label{a7}
\eeq
\beq
C^{+}(L(t,z)-kI)=0\,,
\label{a8}
\eeq
\beq
-\partial_t C^{+}+C^{+}{M}^{(+)}(t,z)=0\,,
\label{a9}
\eeq
where $C^+ =(c_{i}^{+})$,  $L$ is given by (\ref{LaxL}), and
$M^{(+)}$  has the form
(\ref{LaxM}) with changing $\lambda _{i} (t)$ to
\beq
\lambda ^{+}_i (t)=-{\dot{c}^{+}_i\over c^{+}_i}+
(\zeta(\eta)-A)\dot{x}_i -\sum_{j\neq i}(b_j^+a_i)
\Phi (x_j-x_i ,z){c^{+}_j\over c^{+}_i}.
\label{a10}
\eeq
Moreover, it follows that the covector $b_{i}^{+}$ satisfies the equations
\beq
\dot{b}_i^+=-\sum_{j\neq i}b_j^+(b_i^+ a_j)V(x_j-x_i)+ {\lambda}_i^+
b_i^+ \,.
\label{a11}
\eeq
The assumption of the theorem imply that the equations
(\ref{a3}), (\ref{a6}) and
(\ref{a8}), (\ref{a9}) have $N$ linear independent solutions (corresponding
to different values of $k$). The compatibility conditions of these equations
have the form of the Lax equations
\beq
\dot{L}+[M,L]=0 , \;\;\;\; \dot{L}+[M^{(+)}, L]=0\,.
\label{a12}
\eeq
In particular, they imply that $\lambda_i=\lambda_i^+$.

The function $\Phi(x,z)$ satisfies the following functional relations:
\beq
\Phi (x-\eta ,z)\Phi (y,z)-\Phi(x,z)\Phi(y-\eta ,z)=
\Phi(x+y-\eta ,z)(V(-x)-V(-y)),
\label{id1}
\eeq
\beq
\Phi ' (x-\eta ,z)=-\Phi (x-\eta ,z)(V(-x)+\zeta (\eta )-A)-
\Phi (-\eta ,z) \Phi (x,z)
\label{id2}
\eeq
(the constant $A$ is defined in (\ref{1.12})). The first relation
is equivalent to the three-term functional equation for the
Weierstrass $\sigma$-function. The second relation follows from the
first one in the limit $y\rightarrow 0$.

The identities (\ref{id1}) and (\ref{id2}) allow one to prove by direct
computations the following lemma completing the proof:
\begin{lem}
For the matrices $L$ and $M$ defined by (\ref{LaxL}) and
(\ref{LaxM}) respectively, where $a_i$ and $b_i^+$ satisfy the
equations (\ref{1.8}) and the constraints
(\ref{27}), the Lax equations
(\ref{a12}) hold if and only if
$x_i(t)$ satisfy the equations (\ref{23}).
\end{lem}
\square

\noindent{\it Proof of Theorem 2.2.} As it was proved above,
eqs. (\ref{1.6}), (\ref{1.6a}) have $N$ linearly independent solutions if
eqs. (\ref{1.8}) and (\ref{1.9}), constraints (\ref{2.7}) and
Lax equations (\ref{a12}) are fulfilled for some value of the spectral
parameter $z$. But according to the statement of Lemma 2.1,
the Lax equations are then fulfilled for all values of the spectral
parameter $z$. Therefore, for each value of $z$ we can define
the double-Bloch solutions of eq. (\ref{1.6})
with the help of formulas (\ref{1.7}) and (\ref{a1}),
where $c_i$ are components of the common solution of eqs.
(\ref{a3}) and (\ref{a6}).

As it follows from (\ref{a3}),
all addmissible pairs of the spectral
parameters $z$ and $k$ satisfy the characteristic equation
$$R(k,z)\equiv \det\,(kI-L(t,z))=0.$$
At the begining of the next section we prove that this equation defines
an algebraic curve $\hat{\G}$ of finite genus. That will complete the proof
of the theorem.
\square

\noindent{\bf Remark 1.} In the abelian case ($l=1$) the operators of the
Lax pair
by a "gauge" transformation with the matrix $U_{ij}=a_i \delta_{ij}$
can be represented in the form
\beq
L^{(l=1)}_{ij}=\dot x_{i} \Phi(x_i -x_j -\eta , z), \label{LaxL1}
\eeq
\beq
M^{(l=1)}_{ij}=\left( (\zeta(\eta )-A)\dot x_{i}
-\sum_{s \ne i} V(x_i -x_s )\dot{x}_s \right) \delta _{ij} -
(1-\delta _{ij})\dot x_{i}\Phi (x_i -x_j ,z). \label{LaxM1}
\eeq
These formulas are equivalent to the Lax pair obtained in \cite{Ruij2},
\cite{Cal}.

\noindent{\bf Remark 2.} In the abelian case it is ehough to require that
only one of eqs. (\ref{1.6}), (\ref{1.6a}) has $N$ linearly
independent double-Bloch solutions with Bloch multipliers satisfying
conditions (\ref{g5}).
\bigskip

\section{The direct problem}

As it follows from the Lax equation (\ref{a12}), the coefficients of
the characteristic equation
\beq
R(k,z)\equiv \det\,(kI-L(t,z))=0 \label{2.1}
\eeq
do not depend on time. Note that they are invariant with respect
to the symmetries (\ref{26}), (\ref{28}).

\begin{th}
The coefficients $r_i(z)$ of the characteristic equation (\ref{2.1})
\beq
R(k,z)=k^N+\sum_{i=1}^N r_i(z)k^{N-i} \label{2.2}
\eeq
do not depend on $t$ and have the form
\beq
r_i(z)=\phi_i(z) (I_{i,0}+(1-\delta_{l,1})I_{i,1}{\tilde s}_i(z)+
\sum_{s=2}^{m_i} I_{i,s} \p_z^{s-2} \wp(z+\eta)) ,\label{2.3}
\eeq
where
\beq
m_i=i-1, \ \ i=1,\ldots,l; \ \ \ m_i=l-1,\ \ i=l+1,\ldots,N; \label{2.3a}
\eeq
\beq
\phi_i(z)={\sigma(z+\eta)^{(i-2)/2}\sigma(z-(i-1)\eta)
\over \sigma(z-\eta)^{i/2}}. \label{2.3b}
\eeq
\beq
{\tilde s}_i(z)={\sigma(z-\eta)\sigma(z-(i-3)\eta)\over
\sigma(z+\eta)\sigma(z-(i-1)\eta)},  \label{2.3c}
\eeq
In a neighbourhood of $z=-\eta$ the function $R(k,z)$ can be
represented in the form
\beq
R(k,z)=\prod_{i=1}^l(k+(z+\eta)^{-1/2}h_i (z+\eta))
\prod_{i=l+1}^N(k+(z+\eta)^{1/2}h_i (z+\eta)),   \label{2.5}
\eeq
where $h_i(z)$ are regular functions of $z$ in a neighbourhood of $z=0$.
\end{th}
\noindent{\it Proof.} Due to (\ref{1.10}) the matrix elements
$L_{ij}$ are double-periodic
functions of $z$. Therefore, we can consider them as single-valued functions
on the original elliptic curve $\Gamma_0$ with a branch cut between the
points $z=\pm \eta$. First of all let us show that the coefficients
$r_i(z)$ of the characteristic polynomial (\ref{2.1}) are meromorphic
functions on the Riemann surface $\hat {\Gamma}_0$ of the function
$E(z)$ defined by (\ref{1.5c}). (This means that $r_i(z)$ are double-valued
functions of $z$ with square root branching at the points
$z=\pm \eta$.)

This fact can be immediately seen from the "gauge equivalent"
form of $L(t,z)$:
\beq L(t,z)=G(t,z)\tilde L(t,z) G^{-1}(t,z),
\ G_{ij}=\delta_{ij}\left[{\sigma (z-\eta)\over \sigma (z+\eta)}
\right]^{x_i(t)/2\eta} , \label{2.6}
\eeq
where the matrix elements of $\tilde L(t,z)$ have square root branching
at the points
$z=\pm \eta$. Moreover, writing them explicitly,
\beq \tilde{L}_{ij}(t,z)={(b_i^+a_j)\over [\sigma (z-\eta)
\sigma (z+\eta)]^{1/2}}
{\sigma (z+x_i-x_j)\over \sigma(x_i-x_j-\eta)},  \label{2.7}
\eeq
we conclude that $r_{2i}(z)$ are single-valued meromorphic functions
of $z$ (i.e., elliptic functions). Further,
$r_{2i+1}(z)$ are meromorphic functions
on $\hat {\Gamma}_0$, which are odd with respect to the involution
$\hat{\tau}_0: \hat {\Gamma}_0 \to \hat {\Gamma}_0$, interchanging sheets
of the covering
 $\hat {\Gamma}_0 \to \Gamma_0$
(this involution corresponds to changing sign of the square
root: $E(z)\to -E(z)$).
Thus the curve $\hat{\Gamma}$ is invariant with respect to the involution
\beq
\hat{\tau}: \hat{\Gamma} \longmapsto \hat{\Gamma}, \ \
\hat{\tau} (k,E)\longmapsto (-k,-E), \label{2.8}
\eeq
which covers the involution $\hat {\tau}_0$.

Let us consider the factor-curve:
\beq
\Gamma:= \{\hat{\Gamma} /\hat{\tau}\}. \label{2.9}
\eeq
This curve is a $N$-fold ramified covering of the initial elliptic curve,
\beq
\Gamma\longmapsto \Gamma_0\,. \label{2.10}
\eeq
 It can be defined by the equation
\beq
\hat R(K,z)=K^N+\sum_{i=1}^N R_i(z)K^{N-i}=0, \label{2.10a}
\eeq
where
\beq  K=k \left[{\sigma (z-\eta)\over \sigma (z+\eta)}\right]^{1/2},\ \
R_i(z)=r_i(z)\left[{\sigma (z-\eta)\over \sigma (z+\eta)}\right]^{i/2}.
\label{2.10b} \eeq
Let us give some more comments on this statement.
The coefficients $R_j(z)$ of the equation (\ref{2.10a}) are meromorphic
functions of the complex variable $z$ obeying the following monodromy
properties:
\beq
R_j(z+2\omega_{\alpha})=R_j(z)e^{-2j\zeta(\omega_{\alpha})\eta} .\label{2.10c}
\eeq
Eq. (\ref{2.10a}) defines a Riemann surface $\tilde \G$, which is a
$N$-fold covering of the complex plane. Due to (\ref{2.10c})
this surface is invariant with respect to the transformations
\beq z\longmapsto z+2\omega_{\alpha}, \  \ K\longmapsto K
e^{-2\zeta(\omega_{\alpha})\eta}.\label{2.10d}
\eeq
The corresponding factor-surface is an algebraic curve
$\G$, which is the covering of the elliptic curve with periods
$2\omega_{\alpha}$.

Now we switch to (\ref{2.5}). This equality follows from the fact that
the leading term of
$\tilde L(t,z)$ in a neighbourhood of the point $z=-\eta$,
\beq
\tilde{L}_{ij}(t,z)={(b_i^+a_j)\over [\sigma(-2\eta) (z+\eta)]^{1/2}}+
O((z+\eta)^{1/2}), \label{2.11}
\eeq
has rank $l$. The corresponding $N-l$-dimensional subspace of eigenvectors
$C=(c_1,\ldots,c_N)$ with zero eigenvalue
is defined by the equations
\beq \sum_{j=1}^N c_j a_{j,\alpha}=0 , \ \ \ \alpha=1,\ldots,l. \label{2.12}
\eeq

Let us turn to determination of the coefficients in the characteristic
equation (\ref{2.1}). Since the matrix elements of
$\tilde L(t,z)$ have simple poles at the point $z=\eta$, then in case
of general position the function $r_i(z)$ has pole of
order $i$ at this point.
It follows from (\ref{2.5}) that at the point $z=-\eta$ the
function $r_i$, $i=1,\ldots,l$, has pole of order
$i$. If $i=l+1,\ldots,2l$ it has pole of order
$2l-i$. At last, if $i=2l+1,\ldots,N$ it has zero of order $i-2l$, i.e.,
\beq
r_i(z)=(z+\eta)^{-i/2} \rho_i(z+\eta), \ \ \ i=1,\ldots,l, \label{2.13}
\eeq
\beq
r_i(z)=(z+\eta)^{i/2-l} \rho_i(z+\eta), \ \ \ i=l+1,\ldots, N, \label{2.14}
\eeq
where $\rho_i(z)$ are regular functions. All these properties of the
functions $r_i(z)$ allow one to represent them in the form
(\ref{2.3}, \ref{2.3a}). (Note that the function $\phi _{i}(z)$
defined by eq. (\ref{2.3b}) has pole of order $i$ at the point $z=\eta$
and zero of order $i-2$ at the point $z=-\eta$.) \square

\noindent{\bf Important remark.}
It is necessary to emphasize that (\ref{2.5}) implies that the characteristic
equation (\ref{2.1}) defines a singular algebraic curve. Indeed, (\ref{2.5})
implies that $(N-l)$ sheets of the corresponding ramified covering
intersect at the point $(z=-\eta,\, k=0)$. We keep the same notation $\G$
for the algebraic curve with the resolved singularity at this point.

The coefficients $I_{i,s}$ in (\ref{2.3}) are integrals of motion.
The total number of them is equal to
$Nl-l(l-1)/2$ which is exactly half the dimension of the reduced phase
space. It follows from the results of the next section that they are
independent.

\begin{lem} In general position genus $g$ of the spectral
curve $\Gamma$ (defined by eq.
(\ref{2.10a})) is equal to $Nl-l(l+1)/2+1$.
\end{lem}
\noindent{\it Proof.} First let us determine genus $\hat g$ of the
curve $\hat{\Gamma}$
defined by eq. (\ref{2.1}). By the Riemann-Hurwitz formula  $\hat g$
we have $2\hat g-2=2N+\nu$, where $\nu$ is the number of branch points
of $\hat{\Gamma}$ over $\hat{\Gamma}_0$, i.e., the number of
values of $z$ for which
$R(k,z)=0$ has a double root.
This is equal to the number of zeros of $\p_k R(k,z)$
on the surface $R(k,z)=0$ outside the points located above the
point $z=-\eta$ (due to the singularity of the initial
curve mentioned above).
The function $\p_k R(k,z)$ has poles of order $N-1$ above the point $z=\eta$
It has also poles of the same order in $l$ points located above
the point $z=-\eta$ that correspond to first $l$ factors in
(\ref{2.5}). In the other $N-l$ points above
$z=-\eta$ it has zeros of order $(N-l)(N-2l-1)$.
Therefore, $\nu=4lN-2l(l+1)$. The curve $\hat{\Gamma}$ is the two-fold
branched covering of the spectral curve $\Gamma$, the number of branch
points being equal to $2N$ (they are located above $z=\pm \eta$).
The Riemann-Hurwitz formula gives then the relation $2\hat g-2=2(2g-2)+2N$,
which proves the lemma.
\square

The characteristic equation (\ref{2.1}) allows us to define two sets of
distinguished points on the spectral curve (\ref{2.10a}).
It follows from the factorization of $R(k,z)$ (\ref{2.5}) that
the function $k$ has poles at $l$ points
lying on the different sheets over the point $z=-\eta$ (they correspond to
the first $l$ factors in (\ref{2.5})). Let us denote them by
$P_i^+,\ i=1,\ldots,l,$. Since a meromorphic function has as many zeros as
poles, we conclude from
(\ref{2.10b}) and
(\ref{2.5}) that the function $k$ on the unreduced spectral curve
$\hat{\G}$ has $2l$ zeros which do not lie above the point $z=-\eta$.
These zeros correspond to $l$ points $P_i^-, i=1,\ldots,l,$ on the
spectral curve $\G$:  \beq k(P_i^-)=0. \label{2.140} \eeq
In general position there is such a point $P_i^-$ above each zero
$z_i^-$ of the function $r_N(z)$ different from its apparent zero
$z=-\eta$:  \beq r_N(z)=\tilde I_{N,0}{\sigma^{(N-l)/2}
(z+\eta)\over \sigma^{N/2} (z-\eta)} \prod_{i=1}^l \sigma (z-z_i^-).
\label{2.14a} \eeq
In the abelian case ($l=1$) the second marked point $P_1^-$
lies above the point $z=(N-1)\eta$.

\begin{th} The components $\Psi_{\alpha}(x,t,P)$ of the
solution $\Psi(x,t,P)$ to the equation
(\ref{1.6}) are defined on the $N$-fold covering $\G$ of the
initial elliptic curve cutted between the points $P_i^+$ and $P_i^-,\
i=1,\ldots,l$. Outside these branch cuts they are meromorphic.
For general initial conditions the curve $\G$ is smooth. Its genus
equals $g=Nl-{l(l+1)\over 2}+1$ and $\Psi_{\alpha}$ have $(g-1)$ poles
$\gamma_1,\ldots,\gamma_{g-1}$ which do not depend on the
variables $x,t$. In a neighbourhood of
$P_i^+, \ i=1,\ldots,l$ the function
$\Psi_{\alpha}$ has the form
\beq \Psi_{\alpha}(x,t,P)= (\chi_0^{\alpha
i}+\sum_{s=1}^{\infty}\chi_s^{\alpha i}(x,t)(z+\eta)^{s})
(\kappa_i(z+\eta)^{-1})^{x/\eta} e^{\kappa_i(z+\eta)^{-1}t}~\Psi_1(0,0,P),
\label{2.15}
\eeq
where $\chi_0^{\alpha i}$ are constants independent of $x,t$,
and $\kappa_i$ are non-zero eigenvalues of the matrix $(b_i^+a_j)$.
In a neighbourhood of $P_i^-$ the function $\Psi_{\alpha}$ has the form
\beq
\Psi_{\alpha}(x,t,P)=(z-z_i^-)^{x/\eta}~
(\sum_{s=0}^{\infty}\tilde{\chi}_s^{\alpha i}
(x,t)(z-z_i^-)^{s}) ~\Psi_1(0,0,P)
\label{2.16}
\eeq
($z_i^-$ are projections of the points $P_i^-$ to the initial elliptic
curve; they are defined by eq. (\ref{2.14a})). The boundary values
$\Psi_{\alpha}^{(\pm)}$ of the function $\Psi_{\alpha}$ at the opposite
sides of the cuts are connected by the relation
\beq
\Psi_{\alpha}^{(+)}=\Psi_{\alpha}^{(-)}e^{2\pi i x/\eta} . \label{2.17}
\eeq
\end{th}
\noindent{\it Proof.} We start with analitic properties of the
eigenvectors of the Lax matrix.

By $\hat{\Gamma}^*$ denote the curve $\hat{\G}$ with cuts between
the pre-images $P_i^+$ of the point $z=-\eta$ and the pre-images
 $Q_i^-$ of the point
$z=\eta$, $i=1,\ldots,N$.  For a generic point $\hat P$ of the curve
$\hat {\G}$, i.e., for the pair $(k,z)=\hat P$, which satisfies
the equation (\ref{2.1}), there exists a unique eigenvector
 $C(0,\hat P)$ of the matrix
$L(0,z)$ normalized by the condition
$c_1(0,P)=1$.  All other components $c_i(0,\hat P)$ are given by
$\Delta_i (0,\hat P)/ \Delta_1(0,\hat P)$, where $\Delta_i (0,P)$ are
suitable minors of the matrix $kI-L(0,z)$. Thus they are meromorphic
on $\hat {\G}^*$. The poles of $c_i(0,\hat P)$ are the zeros on $\hat
{\G}^*$ of the first principal minor \beq \Delta_1(0,\hat
P)=\det(k\delta_{ij}-L_{ij}(0,z))=0,\ \ i,j>1. \label{2.18} \eeq
Therefore, these poles only depend on the initial data.

\begin{lem}
The coordinates $c_j(0,\hat P)$ of the eigenvector $C(0,\hat P)$ are
meromorphic functions on
$\hat{\G}^*$.
The boundary values $c_j^{\pm}$ of the functions $c_j(0,\hat P)$
at opposite sides of the cuts satisfy the relation
\beq
c_j^+=c_j^- e^{\pi i (x_j(0)-x_1(0))/\eta}. \label{2.19}
\eeq
In a neighbourhood of the point $P_i^+$ the functions $c_j(0, \hat P)$
have the form
\beq
c_j(0,\hat P)=(c_j^{(i,+)}(0)+O(z+\eta)) (z+\eta)^{(x_1(0)-x_j(0))/2\eta},
\label{2.20}
\eeq
where $c_j^{(i,+)}(t)$ are eigenvalues of the residue of the
matrix $\tilde L(0,z)$
at $z=-\eta$, i.e.,
\beq
\sum_{j=1}^N (b_k^+a_j) c_j^{(i,+)}(t)=-\kappa_i c_k^{(i,+)}(t). \label{2.21}
\eeq
In a neighbourhood of the point $Q_i^-$ the functions $c_j(0, \hat P)$
have the form
\beq
c_j(0,\hat P)=(c_j^{(i,-)}(0)+O(z-\eta)) (z-\eta)^{(x_j(0)-x_1(0))/2\eta},
\label{2.22}
\eeq
where $c_j^{(i,-)}(t)$ are eigenvectors of the residue of the
matrix $\tilde L(0,z)$
at $z=\eta$.
\end{lem}
The proof follows from the representation of $C(0,\hat P)$
in the form (see (\ref{2.6})):
\beq
C(0,\hat P)=G(0,z)\tilde C(0,P), \label{2.23}
\eeq
where $G(t,z)$ is defined in (\ref{2.6}), and $\tilde C(0,P)$ is
the eigenvector of the matrix $\tilde L(0,z)$. Matrix elements of
$\tilde L(0,z)$
are analitic on the cuts between $z=\pm \eta$. Hence $\tilde
C(0,P)$ has no discontinuity on the cuts. This proves
(\ref{2.19}). The equations (\ref{2.20}) and (\ref{2.22}) are
direct consequences of the fact that $\tilde L(0,z)$ has simple poles
at the points $z=\pm \eta$.

\noindent{\bf Remark.} The vector
$\tilde C(0,P)$ is invariant under the involution (\ref{2.8})
(that's why its argument is a point $P$ of the spectral curve $\G$
rather than a point $\hat P\in \hat{\G}$). However, this notation is
somewhat misleading since both factors in (\ref{2.23}) are multi-valued
on $\hat{\G}$,
and only their product is well-defined.

\begin{lem}
The poles of $C(0,\hat P)$ are invariant under the involution $\hat {\tau}$.
The number of them is equal to $2Nl-l(l+1)$.
\end{lem}
To prove the lemma we use the following standard argument.
Consider the function of the complex variable $z$ defined by:
$$F(z)=\left( {\rm Det}\left| c_i(0,M_j) \right| \right)^2, $$
where $M_j,\ j=1,\ldots,N,$ are the points above $z$. It is well defined
as a function of $z$ since it does not depend on the ordering of the $M_j$'s.
The analitic properties of $c_j$ allow us to represent $F$ in the form
\beq F(z)=\tilde F(z)\left[\sigma (z-\eta)\over \sigma
(z+\eta) \right]^ {\sum (x_i(0)-x_1(0))},  \label{2.24} \eeq where $\tilde
F$ is a meromorphic function. This means that
$F$ has as many zeros as poles. The number of its poles is twice
the number of zeros of the vector
 $C(0,\hat P)$ whereas the number of zeros of $F$ is equal to the number
 of branch points $\nu$ of the covering
$\hat{\G}$ over $\hat{\G}_0$
(\ref{2.1}).
In the proof of Lemma 3.1 we showed that $\nu=4Nl-2l(l+1)$.
The invariance of poles of $C(0,\hat P)$ under
the involution $\hat {\tau}$ follows from the $\hat {\tau}$-invariance
of eq. (\ref{2.18}) which determines positions of the poles.
This completes the proof.

Let $\gamma_1,\ldots, \gamma_{g-1}$ be the points of the spectral curve
$\G$ whose pre-images are poles of $C(0,\hat P)$. Note that if
$\G$ is smooth, then $g=Nl-l(l+1)/2+1$ coincides with its genus.

Let $C(t, \hat P)$ be the vector obtained from $C(0,\hat P)$
by the time evolution according to eq. (\ref{a6}).

\begin{lem}
The coordinates $c_j(t,\hat P)$ of the vector $C(t,\hat P)$
are meromorphic on $\hat{\G}^*$. Their poles are located above the
points $\gamma_1,\ldots, \gamma_{g-1}$ and do not depend on $t$.
The boundary values $c_j^{\pm}$ of $c_j(t,\hat P)$ at opposite sides of
the cuts satisfy the relation
\beq
c_j^+=c_j^- e^{\pi i (x_j(t)-x_1(0))/\eta}. \label{2.25}
\eeq
In a neighbourhood of $P_i^+$ the functions $c_j(t, \hat P)$  have
the form
\beq
c_j(0,\hat P)=(c_j^{(i,+)}(t)+O(z+\eta)) (z+\eta)^{(x_1(0)-x_j(t))/2\eta}
\exp (\kappa_i(z+\eta)^{-1}t), \label{2.26}
\eeq
where $\kappa_i$ and $c_j^{(i,+)}(t)$ are defined in (\ref{2.21}).
In a neighbourhood of $Q_i^-$ the functions $c_j(t, \hat P)$
have the form
\beq
c_j(t,\hat P)=(c_j^{(i,-)}(t)+O(z-\eta)) (z-\eta)^{(x_j(t)-x_1(0))/2\eta}.
\label{2.26a}
\eeq
\end{lem}
\noindent{\it Proof.}
The fundamental matrix of solutions $S(t,z)$ to the equation
\beq
(\p_t+M(t,z)) S(t,z)=0, \ \  S(0,z)=1, \label{2.27}
\eeq
is a holomorphic function of the variable $z$ outside the cut
connecting the points $z=\pm
\eta$.  We have $L(t,z)= S(t,z) L(0,z)
S^{-1}(t,z)$.  Therefore, the vector $C(t,z)$ equals $C(t,z)=
S(t,z)C(0,z)$ hence it has the same poles as $C(0,P)$.

Let us consider the vector $\tilde C(t,\hat P)$ such that
\beq
C(t,\hat P)=G(t,z)\tilde C(t,\hat P), \label{2.28}
\eeq
where $G(t,z)$ is the same diagonal matrix as in (\ref{2.6}).
This vector is an eigenvector of the matrix $\tilde L(t,z)$ and
satisfies the equation
\beq (\p_t+\tilde M(t,z))\tilde C(t,P)=0,\ \ \tilde
M=G^{-1}\p_tG+G^{-1}MG.  \label{2.29} \eeq
The matrix elements of $\tilde M$
are analitic at the cuts between $z=\pm \eta$. Thus
$\tilde C(t, \hat P)$ is analitic at the cuts on $\hat{\G}$.
Therefore, the multi-valuedness of $C(t, \hat P)$ is fully caused by
the multi-valuedness of $G(t,z)$. This proves eq. (\ref{2.25}).
Eq. (\ref{2.26a}) follows rom the analiticity of $\tilde M$
at the point $z=\eta$. In a neighbourhood of $z=-\eta$ we have:
 \beq
\tilde M_{ij}(t,z)={(b_i^+a_j)\over (z+\eta)}+O((z+\eta)^0). \label{2.30}
\eeq
Therefore, in a neighbourhood of $P_i^+$ it holds
$$\partial_t \tilde C (t,\hat P) = (\mu_i (t,z) +O(z^0))
\tilde C (t,\hat P),$$
where
\beq
\mu_i(t,z)=\kappa_i (z+\eta)^{-1}+O(1), \label{2.31}
\eeq
are eigenvalues of the matrix $\tilde M$. This proves eq. ({\ref{2.26}).
\square

Let us now continue the proof. By the initial definition,
$$\Psi(x,t,\hat P)=\sum_{j=1}^N
s_j(t,\hat P)\Phi(x-x_j(t),z) k^{x/\eta}, \ \ s_j(t,\hat P)=c_j(t,\hat
P)a_j(t),$$
The solutions $\Psi$ of the linear problem (\ref{1.6}) are defined on
the curve $\hat{\G}$. In order to show that $\Psi$ is well defined on the
spectral curve $\G$ we use the equality
\beq
c_j(t,\hat P)\Phi(x-x_j(t),z) k^{x/\eta}=\tilde c_j(t,P)
{\sigma(z+x-x_j+\eta)\over \sigma(z+\eta)\sigma(x-x_j)}
\left[k\left({\sigma(z-\eta)\over \sigma(z+\eta)}\right)^{1/2}\right]^{x/\eta}.
\label{2.32} \eeq

Recall that components of the vector $\tilde C$ are even with respect
to the involution $\hat {\tau}$ (\ref{2.8}). The factor
\beq
K(P)=k\left[{\sigma(z-\eta)\over \sigma(z+\eta)}\right]^{1/2},
\label{2.33}
\eeq
is $\hat {\tau}$-invariant too. Thus
$\Psi(x,t,P)$ is well defined on the spectral curve
$\G$.  At the same time we see that poles of $\Psi(x,t,P)$ coincide
with poles of $\tilde C(0,P)$, i.e., they are located
at the points $\gamma_1, \ldots, \gamma_{g-1}$.

Note that $K(P)$ is a multi-valued meromorphic function on $\G$
with zeros and poles (which are nevertheless well defined)
at the points
$P_i^-$ and $P_i^+,\ i=1,\ldots, l,$  respectively.
Therefore, cutting $\G$ between $P_i^{\pm}, \ i=1,\ldots,l,$ we can choose
the branch of the third factor in (\ref{2.33}) in such a way that
$\Psi$ becomes single-valued outside these cuts, and its boundary
values at the sides of the cuts satisfy the relation (\ref{2.17}).
Consider now the behaviour of $\Psi$
in neighbourhoods of $P_i^+$. In a neighbourhood of $z=-\eta$ we have:
\beq
{\sigma(z+x+\eta)\over \sigma(z+\eta)\sigma(x)}=
{1\over z+\eta}+O(1). \label{2.34}
\eeq
Hence in a neighbourhood of $P_i^+$ it holds
\beq
\Psi_{\alpha}=\sum_{j=1}^N \left({a_{j, \alpha} c_j^{(i,+)}(t)\over z+\eta}+
O(1)\right) \left[k_i(z)\left({\sigma(z-\eta)\over
\sigma(z+\eta)}\right)^{1/2}\right]^{x/\eta},\label{2.35}
\eeq
where $k_i(z)$ is the branch of $k(P)$ defined by $i$-th factor of the
equality (\ref{2.5}). For $i>l$ the product of the second and the third
factors in (\ref{2.35}) is regular in a neighbourhood of $P_i^+$. Since
the eigenvalues $\kappa_i$ in (\ref{2.21}) equal zero for $i>l$, then
the first factor is regular in a neighbourhood of $P_i^+, i>l,$ too.
Therefore, the functions $\Psi_{\alpha}$
are regular at these points.
Similar arguments for $i=1,\ldots, l,$ prove the equality
(\ref{2.15}). Note also that $\Psi_1(0,0,P)$ in (\ref{2.15}) has simple
poles at the points $P_i^+,\ i=1,\ldots,l$.

Consider now the term $\chi_0^{\alpha i}$ in
(\ref{2.15}). By construction, it does not depend on $x$.
Substituting the series (\ref{2.15}) into (\ref{1.6}),
we see that it does not depend on $t$ as well.
\square

The following theorem can be proved by the same arguments.
\begin{th} The components $\Psi^{+,\alpha}(x,t,P)$
of the solution $\Psi^+(x,t,P)$ to the equation
(\ref{1.6a}) are defined on $N$-fold covering $\G$ of the initial
elliptic curve with branch cuts between the points $P_i^+$ and $P_i^-,\
i=1,\ldots,l$. Outside these cuts they are meromorphic. In general
position, the curve $\G$ is a smooth algebraic curve. Its genus equals
$g=Nl-{l(l+1)\over 2}+1$ and $\Psi^{+,\alpha}$ has $(g-1)$ poles
$\gamma_1^+,\ldots,\gamma_{g-1}^+$,
which do not depend on $x,t$. In a neighbourhood of
$P_i^+, \ i=1,\ldots,l$ the function
$\Psi^{+,\alpha}$ has the form
\beq \Psi^{+,\alpha}(x,t,P)= (\chi_0^{+,\alpha
i}+\sum_{s=1}^{\infty}\chi_s^{+,\alpha i}(x,t)(z+\eta)^{s})
(\kappa_i(z+\eta)^{-1})^{-x/\eta}
e^{-\kappa_i(z+\eta)^{-1}t}~\Psi^{+,1}(0,0,P),
\label{2.36}
\eeq
where $\chi_0^{\alpha i}$ are constants independent of $x,t$.
In a neighbourhood of $P_i^-$ the function $\Psi^{+,\alpha}$
has the form
\beq
\Psi^{+,\alpha}(x,t,P)=(z-z_i^-)^{-x/\eta}~
(\sum_{s=0}^{\infty}\tilde{\chi}_s^{\alpha i}(x,t)
(z-z_i^-)^{s}) ~\Psi^{+,1}(0,0,P).
\label{2.37}
\eeq
The boundary values
$\Psi^{+,\alpha;(\pm)}$ of the function $\Psi^{+,\alpha}$ at opposite
sides of the cuts are connected by the relation
\beq
\Psi^{+,\alpha;(+)}=\Psi^{+,\alpha;(-)}e^{-2\pi i x/\eta} . \label{2.38}
\eeq
\end{th}

{\bf Remark.} Theorem 3.2 shows, in particular, that the solution $\Psi$
to the equation (\ref{1.6}) is (up to normalization) the
Baker-Akhiezer function.  In the next section we show that this function
is uniquely defined by the curve $\G$, the poles $\gamma_s$, the matrix
$\chi_0$ and the value $x_1(0)$. All these quantities are defined by
initial Cauchy data and do not depend on $t$. However, it is necessary to
emphasize that part of them depend on the choice of the
normalization point $t_0$,
that we have chosen as ($t_0=0$). Any initial data
$\{x_i,\dot{x}_i,a_i,b_i^+| (b_i^+,a_i)=\dot{x}_i\}$ define the matrix
$L$ by means of eq. (\ref{LaxL}). The characteristic equation
(\ref{2.1}) defines the curve $\G$. Eq. (\ref{2.18}) defines a set of $g-1$
points $\gamma_s$ on $\G$. Therefore, there exists a map
\beq
\{x_i,\dot{x}_i,a_i,b_i^+| (b_i^+,a_i)=\dot x_i\}\longmapsto
\{ \G,\ \ D\in J(\G) \}, \label{2.39}
\eeq
\beq
D=\sum_{s=1}^{g-1}A(\gamma_s)+x_1U^{(0)} ,\label{2.40}
\eeq
where $A:\G\to J(\G)$ is the Abel map and $U^{(1)}$ is a vector
depending on
$\G$ only (see (\ref{3.9b})). The coefficients of the equation
(\ref{2.1}) are integrals of the system (\ref{23})-(\ref{25}).
As we shall see in the next section the second part of the data (\ref{2.39})
define angle-type variables, i.e., the vector $D(t)$ evolves linearly,
$D(t)=D(t_0)+(t-t_0)U^{(+)}$, if
a point of the phase space evolves according to eqs.
(\ref{23})-(\ref{25}).
These equations have the obvious symmetries:
\beq
a_i,b_i^+\to q_i a_i,\ q_i^{-1}b_i^+,\ \ a_i,b_i^+\to
W^{-1}a_i,\ b_i^+W, \label{2.41}
\eeq
where $q_i$ are constants and $W$ is an arbitrary constant matrix.
In the next section we prove that the data $\G,D$ uniquely define
a point of the phase space up the symmetry transformations (\ref{2.41}).

\section{Finite-gap solutions of the non-abelian Toda chain}

Finite-gap solutions of the non-abelian Toda chain were found in the
paper
\cite{kr_ap} by one of the authors. To construct the inverse
spectral transformation for the spin generalizations of the
Ruijsenaars-Schneider model, we recall
the main points of this theory. Since we are working with a continuous
variable $x$ rather than the discrete variable $n$, some minor
modifications of the construction are necessary.

\begin{th}
Let $\G$ be a smooth algebraic curve of genus $g$ with fixed local
coordinates $w_{j,\pm}(P)$ in neighbourhoods of
$2l$ points $P_j^{\pm}, \ w_{j,\pm} (P_j^{\pm})=0, \ j=1,\ldots,l$ and
with fixed cuts between the points $P_j^{\pm}$. Then for each set of
$g+l-1$ points $\gamma_1,\ldots,\gamma_{g+l-1}$ in general position
there exists a unique function $\psi_{\alpha}(x,T,P),\ \alpha=1\ldots,l,
\ T=\{t_{i,j;\pm},\ i=1,\ldots,\infty;\ j=1,\ldots,l \}$ such that

$1^0.$ The function $\psi_{\alpha}$ of the variable $P\in \Gamma$ is
meromorphic outside the cuts and has at most simple poles at the points
$\gamma_s$ (if all of them are distinct);

$2^0.$ The boundary values $\psi^{(\pm)}_{\alpha}$ of this function
at opposite sides of the cuts satisfy the relation
\beq
\psi_{\alpha}^{(+)}(x,T,P)=\psi_{\alpha}^{(-)}(x,T,P)
e^{2\pi ix/\eta}. \label{3.1}
\eeq

$3^0.$ In a neighbourhood of the point $P_j^{\pm}$ it has the form
\beq
\psi_{\alpha}(x,T,P)=w_{j,\pm}^{\mp x/\eta}
\left(\sum_{s=0}^{\infty} \xi_s^{\alpha j;\pm}(x,T)w_{j,\pm}^s \right)
\exp \left(\sum_{i=1}^{\infty}w_{j,\pm}^{-i}t_{i,j;\pm}\right),\ \ \
w_{j,\pm}=w_{j,\pm}(P),\label{3.2}
\eeq
\beq
\xi_0^{\alpha j;+}(x,T)\equiv \delta_{\alpha j}. \label{3.3}
\eeq
\end{th}
The proof of theorems of this kind, as well as the explicit formula
for $\psi _{\alpha}$ in terms of Riemann theta-functions, are
standard in the finite-gap integration theory. We use the notation
of the paper \cite{bab}.

It follows from the Riemann-Roch theorem that for each divisor
$D=\gamma_1+\cdots+\gamma_{g+l-1}$ in general position there exists a
unique meromorphic function $h_{\alpha}(P)$ such that the divisor of its
poles coincides with $D$ and such that
\beq
h_{\alpha}(P_j^+)=\delta_{\alpha j}. \label{3.4}
\eeq
The basis of cycles $a_i^0,\ b_i^0$ on $\G$ with canonical intersection
matrix being fixed, this function may be written as follows:
\beq
h_{\alpha}(P)={f_{\alpha}(P)\over f_{\alpha}(P_{\alpha}^+)}; \quad
f_{\alpha}(P)=\theta(A(P)+Z_{\alpha})
{\prod_{j\neq \alpha}\theta(A(P)+R_j)\over \prod_{i=1}^l\theta
(A(P)+S_i)}, \label{3.5}
\eeq
where the Riemann theta-function
$\theta(z_1,\ldots,z_g)=\theta(z_1,\ldots,z_g|B)$
is defined by the matrix $B=(B_{ik})$ of periods of holomorphic
differentials
on $\G$; $A(P)$ is the Abel map: $A: P\in \G\to J(\G)$;
\beq
R_j=-{\cal K}-A(P_j^+)-\sum_{s=1}^{g-1} A(\gamma_s), \ \ j=1,\ldots,l
\eeq
\beq
S_i=-{\cal K}-A(\gamma_{g-1+i})-\sum_{s=1}^{g-1} A(\gamma_s),
\eeq
and
\beq
Z_{\alpha}=Z_0-A(P_{\alpha}^+), \quad
Z_0=-{\cal K}-\sum_{i=1}^{g+l-1} A(\gamma_i)+\sum_{j=1}^l A(P_j^+),
\label{3.6}
\eeq
where ${\cal K}$ is the vector of Riemann's constants
(the proofs can be found in
\cite{bab}).

Let $d\Omega^{(i,j;\pm)}$ be the unique meromorphic differential
holomorphic on $\G$ outside the point $P_j^{\pm},\
j=1,\ldots,l$, which has the form
\beq
d\Omega^{(i,j;\pm)}=d(w_{j,\pm}^{-i}+O(w_{j;\pm})), \label{3.7}
\eeq
near the point $P_j^{\pm}$ and normalized by the conditions
\beq
\oint_{a_k^0}d\Omega^{(i,j;\pm)}=0. \label{3.8}
\eeq
It defines a vector $U^{(i,j;\pm)}$ with coordinates
\beq U^{(i,j;\pm)}_k={1\over 2\pi i} \oint_{b_k^0} d\Omega^{(i,j;\pm)}.
\label{3.9}
\eeq
Further, define a differential $d\Omega^{(0)}$, which is
holomorphic outside the points $P_j^{\pm}$,
has the form
\beq
d\Omega^{(0)}=\pm
{dw_{j,\pm}\over \eta w_{j,\pm}}+O(1)dw_{j,\pm}, \label{3.9a}
\eeq
near these points, and has zero $a$-periods. It defines a vector
 $U^{(0)}$ with coordinates
\beq U^{(0)}_k={1\over 2\pi i} \oint_{b_k^0} d\Omega^{(0)}
\label{3.9b}
\eeq
It follows from Riemann's bilinear relations that
\beq
U^{(0)}=\eta^{-1} \sum_{j=1}^l (A(P_j^-)-A(P_j^+)). \label{3.9c}
\eeq
\begin{th}
The components $\psi_{\alpha}$ of the Baker-Akhiezer function
$\psi(x,T,P)$ are given by the formula
\beq
\psi_{\alpha}=h_{\alpha}(P) {\theta
(A(P)+U^{(0)}x+\sum_A U^{(A)}t_A+Z_{\alpha}) \theta (Z_0) \over
\theta (A(P)+Z_{\alpha}) \theta (U^{(0)}x+\sum_A U^{(A)}t_A+Z_0) }
e^{(x\Omega^{(0)}(P)+\sum_A t_A\Omega^{(A)}(P))}, \label{3.10}
\eeq
\beq \Omega^{(A)}(P)=\int_{q_0}^P d\Omega^{(A)},\ \ A=(i,j;\pm). \label{3.11}
\eeq
\end{th}
{\bf Remark.} Note that since the abelian integral
$\Omega^{(0)}$ has logarithmic singularities at the marked points,
one may define a single-valued branch of $\psi$ only after cutting
the curve $\G$
between the points $P_j^{\pm}$.

Let us now define the dual Baker-Akhiezer function. For each set of
$g+l-1$ points in general position there exists a unique differential
$d\Omega$ holomorphic outside the points $P_j^{\pm}$ such that it has
simple poles at these points with residues
$\pm 1$, i.e.,
\beq
d\Omega=\pm {dw_{j,\pm}\over w_{j,\pm}}+O(1)dw_{j,\pm} \label{3.12}
\eeq
and such that it equals zero at the points $\gamma_s$:
\beq
d\Omega(\gamma_s)=0. \label{3.13}
\eeq
Becides $\gamma_s$ this differential has $g+l-1$ other zeros which we
denote by $\gamma_s^+$.

The dual Baker-Akhiezer function is the unique function
$\psi^+(x,T,P)$ with components $\psi^{+,\alpha}(x,t,P)$ such that

$1^0.$ The function $\psi^{+,\alpha}$ of the variable $P\in \Gamma$
is meromorphic outside the cuts and has at most simple poles at the
points $\gamma_s^+$ (if all of them are distinct);

$2^0.$ The boundary values $\psi^{+;\alpha,(\pm)}$ of this function
at opposite sides of the cuts satisfy the relation
\beq
\psi^{+;\alpha,(+)}(x,T,P)=\psi^{+;\alpha,(-)}(x,T,P)
e^{-2\pi ix/\eta}; \label{3.1a}
\eeq

$3^0.$ In a neighbourhood of $P_j^{\pm}$ it has the form
\beq
\psi^{+,\alpha}(x,T,P)=w_{j,\pm}^{\pm x/\eta}
\left(\sum_{s=0}^{\infty} \xi_s^{+;\alpha j;\pm}(x,T)w_{j,\pm}^s \right)
\exp \left(-\sum_{i=1}^{\infty}w_{j,\pm}^{-i}t_{i,j;\pm}\right),\label{3.2a}
\eeq
\beq
\xi_0^{+;\alpha j;+}(x,T)\equiv \delta_{\alpha j}. \label{3.3a}
\eeq

Let $h^+_{\alpha}(P)$ be the function that has poles at the points of the
dual divisor $\gamma_1^+,\ldots,\gamma_{g+l-1}^+$ and normalized by
$h_{\alpha}^+(P_j^+)=\delta_{\alpha j}$. It can be written in the form
(\ref{3.5}) in which $\gamma_s$ are replaced by $\gamma_s^+$.
It follows from the definition of dual divisors that
\beq
\sum_{s=1}^{g+l-1} A(\gamma_s)+\sum_{s=1}^{g+l-1} A(\gamma_s^+)=
K_0+\sum_{j=1}^l (A(P_j^+)+A(P_j^-)), \label{3.14}
\eeq
where $K_0$ is the canonical class (i.e. the equivalence class of the
divisor of zeros of a holomorphic differential). Thus the vector $Z_0^+$
in the formulas for $h^+_{\alpha}$ is connected to $Z_0$ by the relation
\beq
Z_0+Z_0^+=-2{\cal K}-K_0+\sum_{j=1}^l (A(P_j^+)-A(P_j^-))=
-2{\cal K}-K_0 - U^{(0)}\eta. \label{3.14a}
\eeq
\begin{th}
The components $\psi^+(x,T,P)$ of the dual Baker-Akhiezer function
are given by
\beq
\psi^{+;\alpha}=h_{\alpha}^+(P) {\theta
(A(P)-U^{(0)}x-\sum_A U^{(A)}t_A+Z_{\alpha}^+) \theta (Z_0^+) \over \theta
(A(P)+Z_{\alpha}^+) \theta (U^{(0)}x+\sum_A U^{(A)}t_A-Z_0^+) }
e^{-(x\Omega^{(0)}(P)+\sum_A t_A\Omega^{(A)}(P))}, \label{3.15}
\eeq
where
\beq
Z_0^+=-Z_0-2{\cal K}-K_0-U^{(0)}\eta , \ \ \
Z_{\alpha}^+=Z_0^+-A(P_{\alpha}^+). \label{3.16}
\eeq
\end{th}

The above results are valid for any algebraic curve with two sets of
marked points. Consider now the class of curves corresponding to the
spin generalizations of the Ruijsenaars-Schneider model.
\begin{th} \label{specific}
Let $\tilde{\G}$ be a smooth algebraic curve defined by an equation
of the form
\beq
\hat R(K,z)=K^N+\sum_{i=1}^N R_i(z)K^{N-i}=0, \label{3.17}
\eeq
where $R_j(z)$ are meromorphic functions of $z$ such that
\beq
R_j(z+2\omega_{\alpha})=R_j(z)e^{-2j\zeta(\omega_{\alpha})\eta}
.\label{3.18} \eeq
and holomorphic in the fundamental domain of the lattice with
periods
$2\omega_{\alpha}$ outside the point $z=-\eta$.
Let us assume that in a neighbourhood of $z=-\eta$ the polynomial
$\hat R$ has the following factorization:
\beq \hat R(K,z)=\prod_{i=1}^l(K+(z+\eta)^{-1}H_i
(z+\eta)) \prod_{i=l+1}^N(K+ (z+\eta)H_i (z+\eta)),   \label{3.19} \eeq
where $H_i(z)$ have no singularity at the point $z=-\eta$.

Then the Baker-Akhiezer function $\psi$ corresponding to: (i)
the curve $\G$, which is the factor of
 $\tilde{\G}$ with respect
 to the transformation group
 \beq z\longmapsto z+2\omega_{\alpha}, \ \
K\longmapsto K e^{-2\zeta(\omega_{\alpha})\eta} , \label{3.20} \eeq
(ii)
local coordinates $w_{j,+}=(z+\eta) H_j^{-1}(0)$ near the poles
$P_j^+,j=1,\ldots,l$ of the multivalued function $K=K(P)$ and arbitrary
local coordinates $w_{j,-}$ near the zeros $P_j^-$ of this function
obeys the relation:
\beq
\psi(x+2\omega_{\alpha},T,P)=\varphi_{\alpha}(P)\psi(x,T,P). \label{3.21}
\eeq
where
\beq
\varphi_{\alpha}(P)=K(P)^{2\omega_{\alpha}/\eta}e^{\zeta(\omega_{\alpha})z}.
\label{3.21a}
\eeq
\end{th}
The proof is easy. It follows from the monodromy properties
(\ref{3.20}) that the values of
$\varphi_{\alpha}(P)$ do not change under shifts of $z$ by periods of
the elliptic curve, i.e., it yields a well-defined function
on $\G$. Eq. (\ref{3.21})
follows from the fact that its left and right hand sides have the
same analitical properties.
\square

\begin{cor}
The Baker-Akhiezer function $\psi(x,T,P)$ with components
$\psi_{\alpha}(x,T,P)$\\
corresponding to the data of Theorem 4.4
can be written in the form
 \beq \psi(x,T,P)=\sum_{i=1}^m
s_i(T,P)\Phi(x-x_i(T)),z)k^{x/\eta},\ \ k=K\left[\sigma(z+\eta)\over
\sigma(z-\eta)\right]^{1/2}. \label{3.22} \eeq
Under these very conditions the dual Baker-Akhiezer function has the form
\beq
\psi^+(x,T,P)=\sum_{i=1}^m s_i^+(T,P)\Phi(-x+x_i(T)-\eta),z)k^{-x/\eta}.
\label{3.22a}
\eeq
\end{cor}
{\it Proof} of the lemma is identical to the proof of the corresponding
lemma from the paper \cite{bab}
and it was already presented at the beginning
of
Sect. 2, because Theorem 4.4 implies that $\psi$ and $\psi^+$ are
double-Bloch functions. (Note that
from (\ref{3.14a}) and (\ref{3.15}) it follows that $\psi^+$
has poles at the points $x_i-\eta$).
\square

So far $t_{i,\alpha;\pm}$ were arbitrary parameters,
entering $\psi$ through the form of the essential singularity
at $P_{\alpha}^{\pm}$. Fix now some values of these parameters
for $i>1$, i.e., $t_{i,\alpha;\pm}=t_{i,\alpha;\pm}^0$, whereas for $i=1$
we put
\beq
t_{1,\alpha;\pm}=t_{\pm}+t_{1,\alpha;\pm}^0 . \label{3.23}
\eeq
The Baker-Akhiezer function now depends on
the variables $(x,t_+,t_-)$. For the sake of brevity, we denote it
by $\psi(x,t_+,t_-,P)$, skipping the dependence on the constants
$T^0$.

\begin{th} For any choice of the constants $T^0$ the Baker-Akhiezer
function \\
$\psi(x,t_+,t_-,P)$
satisfies the equations
\beq
\p_+\psi(x,t_+,t_-,P)=
\psi(x+\eta,t_+,t_-,P) + v(x,t_+,t_-)\psi(x,t_+,t_-,P) ,\label{3.24}
\eeq
\beq
\p_- \psi(x,t_+,t_-,P)=c(x,t_+,t_-)\psi(x-\eta,t_+,t_-,P), \ \
\p_{\pm}=\p/\p t_{\pm}, \label{3.25}
\eeq
where
\beq
v(x,t_+,t_-)= \p_+ g(x,t_+,t_-)g^{-1}(x,t_+,t_-)=\xi_1^+(x,t_+,t_-)-
\xi_1^+(x+\eta,t_+,t_-), \label{3.27}
\eeq
\beq
c(x,t_+,t_-)=g(x,t_+,t_-)g^{-1}(x-\eta,t_+,t_-)=\p_-\xi_1^+(x,t_+,t_-),
\label{3.26}
\eeq
and the matrices $g$ and $\xi_1^+$ are defined by the coefficients
of the expansion (\ref{3.2}):
\beq
g^{\alpha,j}(x,t_+,t_-)=\xi_0^{\alpha,j;-} (x,t_+,t_-),\ \
\xi_1^+(x,t_+,t_-)=\{\xi_1^{\alpha j;+} \} \label{3.28}
\eeq
\end{th}
{\it Proof.} Eq. (\ref{3.24}) follows from the fact that the function
$$\p_+\psi_{\alpha}(x,t_+,t_-,P)-\psi_{\alpha}(x+\eta,t_+,t_-,P)$$
has the same analitic properties as $\psi$ except the normalization
\ref{3.3}). Thus we can write it as a linear combination of the
components $\psi_{\beta}$ with coefficients $v^{\alpha \beta}$.
The explicit form of $v^{\alpha \beta}$ follows from comparing
the coefficients in the left and right hand sides of (\ref{3.24})
at the points $P_j^-$. Hence we get the first equality in
(\ref{3.27}). On the other hand, we may expand $\psi$ near the points
$P_j^+$. Then we get the second equality in
(\ref{3.27}). Eqs.
(\ref{3.25}) and (\ref{3.26}) are proved in a similar way.
\square

\begin{cor} The matrix function $g_n(t_+,t_-)=g(n\eta+x_0,t_+,t_-)$,
corresponding (according to the definitions of the Baker-Akhiezer
functions) to the Riemann surface
$\G$ with fixed local coordinates near the marked points
$P_j^{\pm}$ and to the set of points $\gamma_1,\ldots,g+l-1$ is a
solution of the 2D Toda chain equations (\ref{1.1}).
\end{cor}
{\bf Remark.} The dependence of $g_n$ on the variables $t_{i,j},\
i=1,\ldots,\infty,\ j=1,\ldots,l,$ corresponds to higher flows
of the 2D Toda chain hierarchy.

\begin{th} The dual Baker-Akhiezer function satisfies the equations
\beq
-\p_+\psi^+(x,t_+,t_-,P)=
\psi^+(x-\eta,t_+,t_-,P) + \psi^+(x,t_+,t_-,P)v(x,t_+,t_-) ,\label{3.29}
\eeq
\beq
-\p_- \psi^+(x,t_+,t_-,P)=\psi^+(x+\eta,t_+,t_-,P)c(x+\eta,t_+,t_-),
\label{3.30}
\eeq
where $c(x,t_+,t_-),\ \ v(x,t_+,t_-)$
are the same as in (\ref{3.24}, \ref{3.25}).
\end{th}
{\it Proof.} The same arguments as in the proof of Theorem 4.5 show
that $\psi^+$ obeys eqs. (\ref{3.29}),
(\ref{3.30})  with coefficients $v^+$ and $c^+$ given by
\beq
c^+(x,t_+,t_-)=[\xi_0^{+;-}(x+\eta,t_+,t_-)]^{-1}\xi_0^{+;-}(x,t_+,t_-),
\label{3.31}
\eeq
\beq
v^+(x,t_+,t_-)=-[\xi_0^{+;-}(x,t_+,t_-)]^{-1}\p_+\xi_0^{+;-}(x,t_+,t_-),
\label{3.32}
\eeq
where the matrix elements $\xi_0^{+;-}=\{\xi_0^{+;\alpha j;-}\}$
are determined from (\ref{3.2a}). The coincidence of the coefficients
for $\psi$ and $\psi^+$ is easily seen
from the relation
\beq
[\xi_0^{+;-}]^{-1}=\xi_0^-(x,t_+,t_-)=g(x,t_+,t_-), \label{3.33}
\eeq
which follows from the definition of the dual Baker-Akhiezer function.
To prove (\ref{3.33}), consider the differential
$\psi_\alpha \psi^{+\beta}d\Omega$, where $d\Omega$
is the same as in the definition of the dual divisor $\gamma_s^+$.
It is a meromorphic differential on $\G$ with the only poles at
$P_j^{\pm}$. Its residue at $P_j^+$ is
\beq
{\rm res}_{P_j^+}\psi_\alpha \psi^{+\beta}d\Omega=
\delta_{\alpha,j}\delta_{\beta,j}.\label{3.34}
\eeq
Further, the residue of this differential at $P_j^-$ is
\beq
{\rm res}_{P_j^+}\psi_\alpha \psi^{+\beta}d\Omega=
-\xi_0^{\alpha,j;-}\xi_0^{+;\beta,j;-}. \label{3.35}
\eeq
Since the sum of the residues should be zero, then (\ref{3.33}) follows
 from (\ref{3.34}) and (\ref{3.35}).
\square

\begin{th} Let the curve $\G$, the marked points $P_j^{\pm}$ and local
coordinates in their neighbourhoods be the same as in Theorem 4.4,
then the corresponding algebraic-geometrical "potentials"
$v$ and $c$ in eqs. (\ref{3.24},
\ref{3.25}) are elliptic functions in $x$.
In general position they have the form
\beq
v(x,T)=\sum_{i=1}^N a_i(T)b_i^+(T)V(x-x_i(T)), \label{3.36}
\eeq
\beq
c(x,T)=\p_-\left(S_0(T)+\sum_{i=1}^N a_i(T)b_i^+(T)\zeta(x-x_i(T)\right),
\label{3.37}
\eeq
where $a_i,\ b_i^+$ are some vectors, and $S_0$ is a matrix functions
that does not depend on $x$.
\end{th}
{\it Proof.}
The potentials in eqs. (\ref{3.24}) and (\ref{3.25}) are
elliptic functions due to (\ref{3.21}).
It follows from (\ref{3.10}) that the poles $x=x_i(T)$ of the
Baker-Akhiezer function correspond to solutions of the equation
\beq
\theta(U^{(0)}x+\sum_A U^{(A)}t_A+Z_0)=0. \label{3.38}
\eeq
It is clear from (\ref{3.6})
that for corresponding solutions $(x_i(T),\ T)$
the first factor in the numerator of (\ref{3.10}) is zero at
$P=P_\alpha^+$. At the points $P_\beta^+$, $\beta \neq \alpha$ the
function  $h_\alpha(P)$ vanishes. Therefore, the residue
$\psi^0_{\alpha,i}(T,P)$ of the function $\psi_{\alpha}(x,T,P)$
at  $x=x_i(T)$
(as a function of $P$) has the following analitic properties:

$1^0.$ It is a meromorphic function on $\G$ outside the cuts between
the points $P_j^{\pm}$  and has the same poles as $\psi$;

$2^0.$ Its boundary values
 $\psi^{0;(\pm)}_{\alpha,i}(T,P)$ at opposite sides of the cuts
satisfy the relation
\beq
\psi^{0;(+)}_{\alpha,j}(T,P)=\psi^{0;(-)}_{\alpha,j}(T,P)
e^{2\pi i x_j(T)/\eta}; \label{3.39}
\eeq

$3^0.$ In a neighbourhood of $P_j^{\pm}$ we have
\beq
\psi^0_{\alpha,i}(T,P)=w_{j,\pm}^{\mp x_i(T)/\eta}
\exp(\sum_{s=1}^{\infty}w_{j,\pm}^{-s}t_{s,j;\pm})
F_{i,j,\alpha}^{\pm}(w_{j,\pm}),
\label{3.40}
\eeq
where $F_{i,j,\alpha}^{\pm}$ are regular in these neighbourhoods, and
 \beq
F_{i,j,\alpha}^+(0)=0. \label{3.41}
\eeq
Hence we see that $\psi^0_{\alpha,i}$ has the same analitic properties
as the Baker-Akhiezer function except the following. The regular factor
in the expansions of this function near {\it all} the points $P_j^+$
has the vanishing leading term. For general $x,t_A$ there is no
such function. For the special values $(x=x_i(T),T)$ such function
$\psi_{i0}(T,P)$ does exist and is {\it unique} up to a constant
(with respect to $P$) factor (it is uniquely defined in general case when
$x_i(T)$ is a simple root of the equation (\ref{3.38})). So we can
represent $\psi _{\alpha}$ in the form

\beq
\psi_{\alpha}(x,T,P)={\phi_{\alpha}(T)\;\psi_{i0}(T,P)\over x-x_i(T)}+
O((x-x_i(T))^0).  \label{3.42}
\eeq
It follows from the last equality that the residues
$\rho_i(T)$ of the matrix $\xi_1^+(x,T)$ with matrix elements
$\xi_1^{\alpha j;+}(x,T)$
\beq
\xi_1^+(x,T)={\rho_i(T)\over x-x_i(T)}+O((x-x_i(t))^0) \label{3.43}
\eeq
have rank 1. This means that there exist vectors $a_i(T)$ and covectors
$b_i^+(T)$ such that $\rho_i=a_i(T)b_i^+(T)$. We see from (\ref{3.21}) that
the matrix $\xi_1^+$ obeys the following monodromy properties:
\beq
\xi_1^+(x+2\omega_l)=\xi_1^+(x,T)+2\zeta(\omega_l)r, \label{3.44}
\eeq
where $r$ is a constant. It follows from these equations and (\ref{3.43})
that
$\xi_1^+$ can be written in the form  \beq
\xi_1^+=S_0(T)+\sum_{i=1}^N a_i(T)b_i^+(T) \zeta(x-x_i(T)).\label{3.45}
\eeq
Combining this with the second equalities in (\ref{3.27}) and (\ref{3.26}),
we get the desired result.  \square

{\bf Remark.} In the abelian case ($l=1$) there is an
equivalent expression for $c(x)$ as a product of pole factors:
\beq
c(x,T)=\prod_{i=1}^N {\sigma(x-x_i(T)+\eta) \sigma (x-x_i(T)-\eta)\over
\sigma^2(x-x_i(T))} \label{3.37a}
\eeq
(see (\ref{20})). Comparing the pole terms in (\ref{3.37}) and
(\ref{3.37a}), we get the relations
\beq
\p_+\p_- x_i(T)= -\sigma^2(\eta)\prod_{k\neq i}
{\sigma(x_i-x_k+\eta) \sigma (x_i-x_k-\eta)\over
\sigma^2(x_i-x_k)} , \label{3.371}
\eeq
\beq
\p_+\p_-x_i(T)=- \p_+x_i(T)\p_-x_i(T) \sum_{k\neq i} (V(x_i-x_k)-V(x_k-x_i)).
\label{3.372}
\eeq
It is easy to check that if the dynamics with respect to $t_{\pm}$ is
given by the hamiltonians
$\sigma (\pm \eta) H_{\pm} $ (\ref{21a}), then these
relations follow from equations of motion.

The connection between algebraic-geometrical potentials in eqs.
(\ref{3.24}), (\ref{3.25}) corresponding to equivalent divisors
is well known in the theory of finite-gap integration. Let
$D=\gamma_1+\cdots+\gamma_{g+l-1}$ and
$D^{(1)}=\gamma_1^{(1)}+\cdots+\gamma_{g+l-1}^{(1)}$ be two
equivalent divisors. This means that there exist a meromorphic function
 $h(P)$ on
$\Gamma$ such that $D$ is the divisor of its poles and $D^{(1)}$
is the divisor of its zeros.
\begin{cor} The algebraic-geometrical potentials
$v(x,T), \ c(x,T)$ and $v^{(1)}(x,T),\\ c^{(1)}(x,T)$ corresponding
to $\Gamma,P_j^{\pm},w_{j,\pm}(P)$ and equivalent divisors
 $D$ and $D^{(1)}$
are gauge equivalent:
\beq
v^{(1)}(x,T)=Hv(x,T)H^{-1},\ \  c^{(1)}(x,T)=Hc(x,T)H^{-1},
\ \ H^{\alpha j}=h(P_j^+)\delta^{\alpha j}. \label{3.46}
\eeq
\end{cor}

\begin{cor}
In general position the curve $\Gamma$ satisfies the conditions
of Theorem 4.4 if and only if it is the spectral curve
(\ref{2.10a}) of the Lax matrix $L$ defined by eq. (\ref{LaxL}),
in which $x_i, \dot x_i$ are arbitrary constants and the vectors
 $a_i, \
b_i^+$ satisfy the conditions (\ref{27}).  \end{cor}

It follows from Theorem 3.2 that the Baker-Akhiezer function
  $\Psi_\alpha(x,t,P)/\Psi_1(0,0,P)$
is related to the normalized Baker-Akhiezer function
$\psi_\alpha(x,t,P)$ by the formula
\beq
{\Psi_\alpha(x,t,P)\over \Psi_1(0,0,P)}=\sum_\beta \chi_0^{\alpha\beta}
\psi_\beta(x,t,P), \label{3.47}
\eeq
where $\psi(x,t,P)$ is the Baker-Akhiezer function defined in the
beginning of this section. It corresponds to the following values
of the parameters
$T=\{t_{i,j;\pm}\}$:
\beq
t_{1,j;+}=t,  \ \ t_{i,j,\pm}=0, \ (i,j,\pm)\neq
(1,j,+), \label{3.48}
\eeq
The equality (\ref{3.47}) yields the corollary:

\begin{cor}
Let $a_i(t),b_i(t),x_i(t)$ be solutions to the equations of motion
(\ref{23})-(\ref{25}), then
\beq
\sum_{i=1}^N a_i (t) b_i^+(t)V(x-x_i(t)) = \chi_0 v(x,t)\chi_0^{-1},
\label{3.48a}
\eeq
where $v(x,t)=v(x,\,t_+=t,\,t_-=0)$ is the potential
corresponding (according to Theorem 4.5) to the normalized
Baker-Akhiezer function
$\psi(x,t,P)$ constructed from the data (a curve with marked points)
obeying the conditions of Theorem 4.4.
\end{cor}
\begin{cor} The map
\beq a_i(t),\ b_i^+(t), x_i(t) \longmapsto \{\G,\ [D]\}, \label{3.49}
\eeq
where $[D]$ is the equivalence class of the divisor $D$ is an isomorphism
up to the transformations (\ref{2.41}).
\end{cor}
The curve $\G$ does not depend on time, while
$[D]$ depends on the choice of the initial point
$t_0=0$. The follwoing theorem shows that this dependence
 $[D(t_0)]$ is linearized on the Jacobian.

\begin{th} Let $\G$ be a curve that is defined by the
equation (\ref{3.19}) and
$D=\gamma_1,\ldots,\\
\gamma_{g+l-1}$ be a set of points in general position.
Then the formulas
\beq
\theta (U^{(0)}x_i(t)+U^{(+)}t+Z_0)=0,  \ \ U^{(+)}=\sum_jU^{(1,j,+)},
\label{3.50}
\eeq
\beq a_{i, \alpha}(t)=Q_i^{-1}(t)h_{\alpha}(q_0) {\theta
(U^{(0)}x_i(t)+U^{(+)}t+Z_{\alpha}) \over \theta (Z_{\alpha})} , \label{3.51}
\eeq
\beq b_i^{\alpha}(t)=Q_i^{-1}(t)h_{\alpha}^+(q_0)
{\theta (U^{(0)}x_i(t)+U^{(+)}t-Z_{\alpha}^+) \over \theta (Z_{\alpha}^+)} ,
\label{3.52}
\eeq
where
\beq Q_i^2(t)={1\over 2}
\sum_{\alpha=1}^lh_{\alpha}^+(q_0) h_{\alpha}(q_0)
{\theta (U^{(0)}x_i(t)+U^{(+)}t-Z_{\alpha})
\theta (U^{(0)}x_i(t)+U^{(+)}t-Z_{\alpha}^+) \over \theta (Z_{\alpha}) \theta
(Z_{\alpha}^+)},    \label{3.53}
\eeq
($q_0$ is an arbitrary point of the curve $\G$)
define solutions of the system (\ref{23}), (\ref{1.8}), (\ref{1.9}).
Any solution of this system in general position may be obtained from
(\ref{3.50})-(\ref{3.52}) by means of the symmetries (\ref{2.41}).
\end{th}

{\bf Remark.} If in eqs. (\ref{3.50})-(\ref{3.52}) the vector $Z_0$
is substituted by
\beq
Z_0\longmapsto Z_0+\sum_A U^{(A)}t_A , \label{3.54}
\eeq
then the corresponding quantities
 $x_i(T),\ a_i(T),\ b_i(T),\
T=\{t_A\},$ depende on $t_A$ according to the higher commuting flows
of the system (\ref{23})-(\ref{25}). We would like to emphasize that
the points $P_j^{\pm}$ enter symmetrically. This means that the dependence
of $x_i(T),\ a_i(T),\ b_i(T)$ on the variable $t_-=l^{-1}\sum_{j=1}^l
t_{1,j;-} $ is described by the same equations as that for
$t=t_+$.

\section{Difference analogs of Lame operators}

Consider the operator $S_0$ given by eq. (\ref{10}) for integer $\ell$.
First of all we note that due to the obvious symmetry
$-\ell \leftrightarrow \ell -1$ it is enough to consider only positive
values of $\ell$. In what follows we imply that $\ell \in {\bf Z}_{+}$.
The finite-gap property of the operator $S_0$
means that the Bloch solutions to the equation
\beq
(S_0 f)(x)=\varepsilon f(x) \label{4.1}
\eeq
are parametrized by points of a hyperelliptic curve of genus $2\ell$.

Any solution $f(x)$ of eq. (\ref{4.1}) may be represented in the form
\beq
f(x)=\Psi(x) \left (\theta _{1}(\eta/2)\right )^{-x/\eta }
\prod_{j=1}^{\ell} \theta_1(x-j\eta), \label{4.2}
\eeq
where $\Psi(x)$ satisfies the equation
\beq
(\tilde S_0 \Psi)(x)\equiv
\Psi(x+\eta)+ c_{\ell}(x)\Psi(x-\eta)=\varepsilon \Psi(x), \label{4.3}
\eeq
\beq
c_{\ell}(x)=\theta_1^2(\eta/2){\theta_1(x+\ell\eta)
\theta_1(x-(\ell +1)\eta)\over \theta_1(x)\theta_1(x-\eta)}. \label{4.4}
\eeq
This transformation sends Bloch solutions of the first equation to
Bloch solutions of the second one. To construct them explicitly, we
use the ansatz similar to the one for the linear equation (\ref{1.2}):
\beq
\Psi=\sum_{j=1}^{\ell} s_j(z,k,\varepsilon)
\Phi_{\ell}(x-j\eta,z) k^{x/\eta}, \label{4.5}
\eeq
where
\beq
\Phi_{\ell}(x,z)={\theta_1(z+x+N\eta)\over \theta_1(z+N\eta) \theta_1(x)}
\left[{\theta_1(z-\eta)\over \theta_1(z+\eta)}\right]^{x/2\eta},
\ \ \ N={\ell (\ell +1)\over 2}. \label{g6}
\eeq
(note that $\Phi_1$ coincides with $\Phi(x,z)$  given by
(\ref{1.5}) and $c_{1}(x)$ coincides with $c(x-\eta)$ from (\ref{1.5b})
up to a constant factor).

The function $\Phi_{\ell}(x,z)$ is double-periodic in $z$:
\beq
\Phi_{\ell}(x,z+2\omega_{\alpha})=\Phi_{\ell}(x,z).  \label{g7}
\eeq
(In this section we take the periods to be $2\omega_1=1,\,
2\omega_2=\tau$.)
For values of $x$ such that $x/2\eta$ is a half-integer number the function
$\Phi_{\ell}$ is single-valued on
the Riemann surface $\hat {\G}_0$ of the function $E(z)$ defined by
(\ref{1.5c}).
For general values of $x$ one can define a single-valued branch of
$\Phi_{\ell}(x,z)$ by cutting the elliptic curve $\G_0$ between the
points $z=\pm \eta$.

As a function of $x$, $\Phi_{\ell}(x,z)$ is a double-Bloch function:
\beq
\Phi_{\ell}(x+2\omega_{\alpha}, z)=
T_{\alpha}^{(\ell)}(z)\Phi_{\ell} (x,
z), \label{g9} \eeq
where
\beq T_1^{(\ell)}(z)=\left (\frac{\theta_1
(z-\eta )}{\theta_1 (z+\eta )}\right )^{1/2\eta}.  \label{g10} \eeq
\beq
T_2^{(\ell )}(z)=\exp (-2\pi i(z+N\eta ))
\left (\frac{\theta _1 (z-\eta )}{\theta _1(z+\eta )}
\right )^{\tau/2\eta}.
\label{g10a}
\eeq
In the fundamental domain of the lattice defined by
$2\omega_{\alpha}$ the function $\Phi_{\ell}(x,z)$ has a unique
pole at the point $x=0$:
\beq
\Phi_{\ell}(x,z)=\frac{1}{\theta_1 '(0)x}+O(1)\,.
\label{g11} \eeq

Substituting (\ref{4.5}) into
(\ref{4.3}) and computing the residues at the points
$z=j\eta,\ j=0,\ldots,\ell$, we get $\ell +1$ linear equations
\beq
\sum_{j=1}^{\ell} L_{i,j} s_j=0,\ \ i=0,\ldots, \ell , \label{4.6}
\eeq
for  $\ell$ unknown parameters $s_j=s_j(z,k,\varepsilon)$.
Matrix elements $L_{i,j}$ of this system are:
\beq
L_{0,1}=k+h\Phi_{\ell}(-2\eta,z)k^{-1},\ \ \
L_{0,j}=h\Phi_{\ell}(-(j+1)\eta,z)k^{-1},\ \ \ j=2,\ldots, \ell ;
\label{4.7}
\eeq
\beq L_{1,1}=-\varepsilon-h\Phi_{\ell}(-\eta,z)k^{-1},\
L_{1,2}=k-h\Phi_{\ell}(-2\eta,z)k^{-1},
\ L_{1,j}=-h\Phi_{\ell}(-j\eta,z)k^{-1},\ \ j>2; \label{4.8}
\eeq
\beq
L_{i,j}=\delta_{i,j+1} c_i k^{-1}-
\varepsilon\delta_{i,j}+\delta_{i,j-1}k,\;\;\;\;\;\; i>1,
\label{4.9}
\eeq
where
\beq
h=\theta_1 '(0){\rm res}_{x=0}c_{\ell}(x)=
\frac{\theta_{1}^{2}(\eta /2)}{\theta_{1}(\eta)}
\theta_1(\ell \eta) \theta_1((\ell +1)\eta),
\label{4.10}
\eeq
\beq
c_j=c_{\ell}(j\eta)=\theta_1^2(\eta /2)
{\theta_1((j+\ell )\eta) \theta_1((j-\ell -1)\eta)\over
\theta_1(j\eta)\theta_1((j-1)\eta)}. \label{4.11}
\eeq
The overdetermined system (\ref{4.6}) has non-trivial solutions if
and only if
rank of the rectangular matrix $L_{i,j}$ is less
than $\ell$. By $L^{(0)}$ and $L^{(1)}$ denote
$\ell \times  \ell$ matrices obtained from $L$ by deleting the rows
with $i=0$ and $i=1$,
respectively. Then the set of parameters
$z,k,\varepsilon$ for which eq.
(\ref{4.3}) has solutions of the form (\ref{4.6}) is defined by the system
of two equations:
\beq \det L^{(i)}\equiv R^{(i)}(z,k,\varepsilon)=0, \ \ i=0,1.
\label{4.12} \eeq
Expanding the determinants with respect to the upper row, we get
that $R^{(i)}$ have the form:
\beq
R^{(0)}(z,k,\varepsilon)=r^{(0)}_{\ell}(\varepsilon)+
\sum_{j=1}^{\ell}k^{-j}\Phi_{\ell}(-j\eta,z)r_{\ell -j}^{(0)}(\varepsilon),
\label{4.13}
\eeq
\beq
R^{(1)}(z,k,\varepsilon)=k\tilde r_{\ell
-1}(\varepsilon)+\sum_{j=1}^{\ell} k^{-j}\Phi_{\ell}(-(j+1)\eta,z)
r_{\ell -j}^{(1)}(\varepsilon), \label{4.13a}
\eeq
where $\tilde r_{\ell -1},\ r_{\ell -j}^{(0)},\ r_{\ell -j}^{(1)}$ are
polynomials in
$\varepsilon$ of degrees $\ell -1$ and $\ell -j$ respectively.

These equations define an algebraic curve $\hat \G$
realized as $\ell (\ell +1)/2$-fold ramified covering of the genus 2 curve
$\hat {\G}_0$ on which the functions
$\Phi_{\ell}(-j\eta,z)$ are single-valued.
This curve enjoyes the obvious symmetry
\beq
(z,\ k,\ \varepsilon)\longmapsto
(z, -k, -\varepsilon) .\label{g13}
\eeq
This is a direct corollary of the following general property
of eq. (\ref{4.3}). Let $\Psi (x)$ be a solution of (\ref{4.3}) with an
eigenvelue $\varepsilon$, then $\Psi(x) \exp(\pi i x/\eta)$ is a solution
of the same equation with the eigenvalue $- \varepsilon$. This
transformation corresponds to the change of sign for $k$.

Note that the transformation
$k \rightarrow -k$ in (\ref{4.5}) and the simultaneous interchanging of
sheets of $E(z)$ leaves the function $\Psi(x,z,k)$ invariant.
Therefore, $\hat {\G}$ may be considered as a $\ell (\ell +1)$-fold
ramified covering of $\G _{0}$.

Let us show that this curve is invariant with respect to another
involution,
\beq
(z,\ k, \ \varepsilon) \longmapsto (-z, \
k^{-1}\theta_1^2(\eta/2), \ \varepsilon),
\label{g14}
\eeq
as well.

Let $\Psi(x)$ be a solution of eq. (\ref{4.3}). Then the function
\beq
\tilde \Psi(x)=\Psi(-x) A(x), \label{g15}
\eeq
where
\beq
A(x)=\theta_1(\eta/2)^{2x/\eta}\prod_{j=1}^{\ell} {\theta_1(x+j\eta)\over
\theta_1(x-j\eta)} \label{g16}
\eeq
is also a solution of the same equation. Moreover, if $\Psi$ is a
double-Bloch solution then $\tilde \Psi$ is also a double-Bloch solution.
It is easy to check that if Bloch multipliers of $\Psi$ are
parametrized by the pair $z, k$,
 then the Bloch multipliers of $\tilde \Psi$
correspond to the pair $-z, \ k^{-1}\theta_1^2(\eta/2)$.

The variable $\varepsilon$ being considered as a function $\varepsilon(P)$
on the curve $\hat \G$ ($P\in \hat \G$) is a meromorphic function. It can
not take any value more than twice because for a given value of
$\varepsilon$ the second order difference equation (\ref{4.3}) has at
most two different Bloch solutions. The involution (\ref{g14}) is not
trivial, so the function $\varepsilon$ does take generic value
two times. Therefore, the algebraic curve $\hat \G$ is a hyperelliptic
curve of finite genus $g$. Due to the symmetry (\ref{g13}) it can be
represented in the form \beq
y^2=\prod_{i=1}^{g+1}(\varepsilon^2-\varepsilon_i^2).\label{g16a}
\eeq
The involution (\ref{g14}) is the
hyperelliptic involution corresponding to the
interchanging of the sheets of the ramified covering (\ref{g16a}).

Now we are going to prove that $g=2\ell$. From (\ref{g16a}) it follows
that there are $2g+2$ fixed points of the hyperelliptic involution.  On
the other hand, the number of fixed points of the hyperelliptic
involution (\ref{g14}) is equal to the number of preimages of the
second-order points $\omega_a \in \G _0,\ a=0,\ldots,3$, (that are fixed
points of the involution $z\to -z$ on $\G_0$) on $\hat \G$ such that the
corresponding value of $k$ is equal to $\pm \theta_1(\eta/2)$.
Explicitly, from now on we adopt the following notation \footnote{this
notation differs from that adopted in Sect.2.}:
\beq
\omega_0=0, \ \omega_1=1/2, \ \omega_2=(1+\tau)/2, \ \omega_3=\tau/2\,.
\label{g201}
\eeq

For values of $\varepsilon$ corresponding to the fixed points
there is only one Bloch solution.
This implies that the corresponding solution has a definite
parity with respect to the involution (\ref{g15}), i.e.
\beq
\Psi(x)=\nu \Psi(-x) A(x), \ \ \ \ \nu=\pm 1. \label{g17}
\eeq
We are going to prove that $\nu=(-1)^{\ell}$.

The equality (\ref{g17}) for $x=\eta$ implies that
\beq
s_1=\nu k^{-1}\Psi(-\eta) (-1)^{\ell -1}
\frac{\theta_{1}^{2}(\eta /2)}{\theta_{1}(\eta)}
\theta_1(\ell \eta) \theta_1((\ell +1)\eta),
\label{g18}
\eeq
Comparing this equality with (\ref{4.3}) (taken at $x=0$), we see that
if $s_1\neq 0$, then $\nu=(-1)^{\ell}$. Otherwise, $s_1=0$ and
$\Psi(-\eta)=0$.

{}From (\ref{4.3}) it follows that coefficients $s_1$ and $s_2$ in
the representation (\ref{4.5})
of a double-Bloch solution can not equal
zero simultaneously. Indeed, let $j$ be a
minimal index such that $s_j\neq 0$.
Suppose $j>2$, then the left hand side of (\ref{4.3}) has a pole at the
point $z=(j-1)\eta$ but the right hand side has no pole at this point.

Therefore, if $s_1=0$, then $s_2\neq 0$. In that case
eq. (\ref{4.3}) implies that $\Psi(0)\neq 0$. At the same time from
(\ref{g17}) (taken at $x=0$) it follows that
\beq
\Psi(0)=(-1)^{\ell}\nu \Psi(0). \label{g20}
\eeq
Therefore, $\nu=(-1)^{\ell}$.

For $z=\omega_a$,
$k=(-1)^{a+1} \theta_1(\eta/2)$ the representation (\ref{4.5})
of the double-Bloch functions is equivalent to
\beq
\Psi(x)=(\theta_1(\eta/2))^{x/\eta}
\exp(\pi i (\delta _{2,a} +\delta _{3,a})x)
\prod_{j=1}^{\ell}\frac{ \theta_1(x+x_j)}
{\theta_1(x-j\eta)},
\label{g21} \eeq where \footnote{cf. the "sum rule" in \cite{Tak1}.}
\beq \sum_{j=1}^{\ell} x_j =
\omega_a\,. \label{g22} \eeq
\begin{lem} The hyperelliptic involution of
$\hat \G$ has $2d$ fixed points, where $d$ equals sum of dimensions of
the functional subspaces consisting of functions that have  the form
(\ref{g21}) and satisfy the relation (\ref{g17}) with
$\nu=(-1)^{\ell}$.
\end{lem}
{\it Proof.} As it was shown above, for each pair of fixed points of
the hyperelliptic involution of $\hat{\Gamma}$ that are invariant
with respect to the involution (\ref{g13}) (corresponding to the change of
sign for $k$) there exists the unique solution to
eq. (\ref{4.3}) having the form (\ref{g21})
and satisfying (\ref{g17}) with $\nu=(-1)^{\ell}$.
On the other hand, the space of
such
functions is invariant with respect to the
operator $\tilde S_0$. Indeed, the
equality (\ref{g18}) with $\nu=(-1)^{\ell}$ (that is a corrolary of
(\ref{g17})) implies that $\tilde S_0 \Psi$ has no pole at $z=0$. At the
same time $\tilde S_0$ commutes with the linear operator (\ref{g15}).
Therefore, the number of solutions of (\ref{4.3}) having  the form
(\ref{g21}) and satisfying (\ref{g17}) is equal to the number of
eigenvalues $\varepsilon_i, \ i=1,\ldots, d$, of $\tilde S_0$ on these
finite-dimensional spaces, i.e., is equal to sum of their dimensions.
\square

It is easy to see that dimension of the space defined
in the last lemma for
$\omega_0=0$ is equal to $(\ell -1)/2$ if $\ell$ is odd and $\ell /2+1$
if $\ell$ is even.  For the other three points of second order
corresponding dimensions are equal to $(\ell +1)/2$ if $\ell$ is odd and
$\ell /2$ if $\ell$ is even. Therefore, the number of fixed points is
$4\ell +2$ and we do prove that $g=2\ell$.

\begin{lem}
The direct sum of spaces of functions having the form (\ref{g21})
and satisfying (\ref{g17}) with $\nu=(-1)^{\ell}$ is invariant with
respect to the operators $\tilde S_a$,
\beq
\!(\tilde S_a \Psi)(x)\!=\!H_a \!
\left[{\theta _{a+1}(x-\ell \eta)\over \theta_1 (x-\ell \eta)}
\Psi (x+\eta)\!-\!
\theta_1^2(\eta/2){\theta_1(x-(\ell +1)\eta)\theta _{a+1}(-x-\ell
\eta)\over \theta_1(x) \theta_1(x-\eta)}\Psi (x-\eta)\right],
\label{g23}
\eeq
\beq
H_a=(i)^{\delta_{a,2}}{\theta_{a+1}(\eta/2)\over \theta_1(\eta/2)}
\label{g23a} \eeq which are gauge equivalent to the operators
(\ref{10}).  \end{lem} The proof is straightforward. Again, eq.
(\ref{g18}) implies that $\tilde S_a \Psi$ has no pole at $x=0$. At the
same time these operators commute with the transformaton (\ref{g15}) and
keep invariant the set of Bloch multipliers corresponding to the spaces
of functions having the form (\ref{g21}). Note that coefficients of
$\tilde S_a, \ a\neq 0$ are not elliptic; that's why they do not
preserve each space but only their direct sum.

\noindent{\bf Remark.}
The invariant space for the operators
$\tilde S_a$ described above coincides
(after the gauge transformation (\ref{4.2})) with the
finite-dimensional representation space of
the Sklyanin algebra found in
\cite{skl2}. More information on invariant functional subspaces for
Sklyanin's operators is contained in the next section.

Equations (\ref{4.12}) represent $k$ and $\varepsilon$ as multi-valued
functions of $z$. Let us consider now analytical properties of
the double-Bloch solutions
$\Psi(x,Q),\  Q=(z,\ k,\ \varepsilon)\in \hat {\G}$
on the hyperelliptic curve $\hat \G$.

\begin{th}
The Bloch solution $\psi(x,P)=\Psi(x,P)\Psi^{-1}(0,P)$
of eq. (\ref{4.3}) is a meromorphic function on the genus
$2\ell$ hyperelliptic
curve $\hat \G$ defined by the equation
\beq
y^2=\prod_{i=1}^{2l+1} (\varepsilon^2-\varepsilon_i^2), \label{g100}
\eeq
(where $\varepsilon_i$ are eigenvalues of $\tilde S_0$ on
the finite-dimensional invariant space of functions having the form
(\ref{g21}), (\ref{g22})) outside a cut between the points
$P^{\pm}$ at infinity (that are preimages of $\varepsilon=\infty$ on
$\hat \G$). Outside the cut $\psi(x,P)$ has $2\ell$ poles independent
of $x$. They are invariant with respect to the involution of $\hat \G$
that covers the involution $\varepsilon \to -\varepsilon$ (\ref{g13}).
The boundary values of $\psi^{(\pm)}(x,P)$ at opposite sides of the cut
are connected by the relation \beq \psi^{(+)}=\psi^{(-)}e^{2\pi i/\eta}.
\label{4.18} \eeq In a neighbourhood of $P^{\pm}$ the function
$\psi(x,P)$ has the form:  \beq \psi=\varepsilon^{\pm x/\eta}
\left(\sum_{s=0}^{\infty} \xi_s^{\pm}(x) \varepsilon^{-s}\right), \ \
\xi_0^+\equiv 1.  \label{4.19} \eeq
\end{th}
{\it Proof.} Coefficients $s_j$ in
(\ref{4.5}) are solutions of the linear
system (\ref{4.6}). Let us normalize them by the condition $s_1=1$.
Then all other $s_j$ are meromorphic functions on $\hat \G$. Therefore,
we may conclude that $\Psi$ (as a function of $Q \in \hat \G$)  is
well defined on $\hat \G$ with cuts between zeros and
poles of
\beq K=\left( {\theta_1(z-\eta)\over
\theta_1(z+\eta)}\right)^{1/2} k(z).  \label{g101} \eeq
At the edges of
these cuts $\Psi$ has a singularity of the form
\beq
\Psi \sim K^{x/\eta} \label{g102} \eeq
(up to a factor of order $O(1)$). From (\ref{4.3}) it directly
follows that a function $\Psi$ with such type of singularity might be a
solution of this equation if
$\varepsilon=\infty$ at the singular points.
On the other hand, in neighbourhoods of
these points we have $K \sim \varepsilon^{\pm 1} $. That proves (\ref{4.19}).
The proof that the number of poles of $\psi$ is equal to $2\ell$ is
standard. First of all note that poles of $\psi$ do not depend on $x$
because poles of $s_j$ do not depend on $x$. Then we may consider the
following meromorphic function of $\varepsilon$:
\beq
F(\varepsilon)=(\psi(\eta,P_1(\varepsilon))-\psi(\eta,P_2(\varepsilon)))^2,
\label{g103}
\eeq
where $P_i(\varepsilon)$ are two preimages of $\varepsilon$ on $\hat \G$.
This function does not depend on the ordering of
these points, i.e., $F$ is a meromorphic function of $\varepsilon$.
It has double poles at projections of poles of $\psi$, has a pole of
order 2 at infinity and has simple zeros at the branch points. The numbers
of zeros and poles of a meromorphic function are equal to each other.
That implies that $\psi$ has $g$ poles.
\square

We would like to mention that the theorem could be proved by a direct
consideration of analytic
properties of $\Psi$ on $\hat \G$ represented in the
form given by (\ref{4.12}). This alternative
proof is very similar to the proof of Theorem 3.2. We skip it, but
present some arguments explaining why the multi-valued function $K$
has only one pole and one zero.

For $\ell>1$ the function
$\Phi_{\ell}(-j\eta,z)$ has a pole and a zero of order
$j$ at the points $z=\pm \eta$, respectively. Therefore, $k$ has zeros and
poles at all preimages of $z=-\eta$ (resp., $z=\eta$) on $\hat \G$.
Hence $K$ is regular at all these points. The function $\Phi_{\ell}$ has
a simple pole at the point $z=-N\eta$. It turns out that at one of the
preimages of this point the function $\varepsilon$ (as well as $k$) has a
pole.
The corresponding point is one of the infinities on
$\hat \G$ (representated
it in the form (\ref{4.12})). The other point at infinity
is one of the preimages of the
point $z=N\eta$.

\noindent{\bf Remark.} Eqs. (\ref{10}) define one particular series
of representations of the Sklyanin algebra found in \cite{skl2}
(called series a) there). In this series the operator
 $S_0$ has the finite-gap property. It should be noted that
 in each of the other series there is again an operator having the
 finite-gap property (see (\ref{5.19}) below).
 The other three operators in each
 series in general do not have this property. Trigonometric
 degenerations of these operators were considered in \cite{GZ}.
 In this case $S_0$ corresponds to solitonic solutions of the Toda chain.
\bigskip

\section{Representations of the Sklyanin algebra}

In this section we show how to construct representations of the
Sklyanin algebra from "vacuum vectors" of the $L$-operator
(\ref{4}). The notions of vacuum vectors and the vacuum curve of
an $L$-operator were introduced by one of the authors \cite{Krichvac}
in analysis of the Yang-Baxter equation by methods of algebraic
geometry. Let us recall the main definitions.

To begin with, consider an {\it arbitrary} $L$-operator ${\cal L}$
with two-dimensional auxiliary space ${\bf C}^2$, i.e., an arbitrary
$2n\times 2n$ matrix represented as $2\times 2$ matrix whose matrix
elements are $n\times n$ matrices ${\cal A}$, ${\cal B}$, ${\cal C}$,
${\cal D}$:
\beq
{\cal L}=\left ( \begin{array}{cc} {\cal A}& {\cal B}\\
{\cal C}& {\cal D} \end{array} \right ).
\label{5.0}
\eeq
The operators ${\cal A},\,{\cal B},\,{\cal C},\, {\cal D}$ act in
a linear space ${\cal H} \cong {\bf C}^n$
which is called the {\it quantum space} of
the $L$-operator. We emphasize that so far no conditions on ${\cal L}$
are implied. In particular, we do not impose the relation (\ref{5})
and do not imply any specific parametrization of the matrix elements.

Let us consider a vector $X\otimes U \in {\cal H}\otimes {\bf C}^2 $
($X\in {\cal H},\, U\in {\bf C}^2 $) such that
\beq
{\cal L}(X\otimes U)=Y\otimes V,
\label{5.1a}
\eeq
where $Y\in {\cal H}$, $V\in {\bf C}^2 $ are some vectors. The relation
(\ref{5.1a}) means that the undecomposable tensor $X\otimes U$
is transformed by ${\cal L}$ into another undecomposable tensor.
Being written in components, this relation has the form
\beq
{\cal L}^{i \alpha}_{j \beta}X_{j}U_{\beta}=Y_{i}V_{\alpha},
\label{5.1b}
\eeq
where indices $\alpha ,\,\beta$ (resp., $i,\,j$) correspond to ${\bf C}^2$
(resp., ${\cal H}$) and summation over repeated indices is implied.

Suppose (\ref{5.1a}) holds; then the vector $X$ is called
a {\it vacuum vector} of the $L$-operator ${\cal L}$. Multiplying
(\ref{5.1a}) from the left by the covector $\tilde {V}=(V_2, -V_1)$,
orthogonal to $V$, we get
\beq
(\tilde {V}{\cal L}U)X=0\,.
\label{5.2}
\eeq
Here $\tilde{V}{\cal L}U$ is an operator in ${\cal H}$ with matrix
elements $\tilde{V}_{\alpha}{\cal L}_{j\beta}^{i\alpha}U_{\beta}$.
Conversely, suppose (\ref{5.2}) holds. Then we have (\ref{5.1a})
with some vector $Y$ which is uniquely determined by $U,\, V$ and $X$.

The relation (\ref{5.1a}) (in the particular case ${\cal H}\cong
{\bf C}^2$) was the starting point for Baxter in his solution of
the eight-vertex model \cite{Baxter}. In the papers on integrable
lattice models of statistical physics this relation is
called "pair propagation through a vertex". In the context of
quantum inverse scattering method \cite{ft} the equivalent condition
(\ref{5.2}) is more customary. It defines the local vacuum of the
gauge-transformed $L$-operator (this explains the terminology
introduced above). In general form this relation appeared for the
first time in \cite{Krichvac}.

The necessary and sufficient condition for the existence of vacuum
vectors is
\beq
\det (\tilde{V}{\cal L}U)=0\,.
\label{5.3a}
\eeq
Putting for simplicity $U_2 =V_2 =1$ and using (\ref{5.0}), one may
represent (\ref{5.3a}) in a more explicit form:
\beq
\det (U_1 {\cal A}+{\cal B}-U_1 V_1 {\cal C}-V_1 {\cal D})=0,\;\;\;\;
	U_2 =V_2 =1.
	\label{5.3b}
	\eeq
This equation defines an algebraic curve in ${\bf C}^2$ which is called
the {\it vacuum curve} of the $L$-operator ${\cal L}$. So vacuum vectors
are parametrized by points of the vacuum curve (i.e. by pairs
$(U_1 , V_1 )$ satisfying (\ref{5.3b})).  In general
position the space of vacuum vectors corresponding to each point of
the curve is one-dimensional.

Suppose now that ${\cal H}\cong {\bf C}^2$ and ${\cal L}$ satisfies
the equation (\ref{5}) with some matrix $R$. In this case the vacuum
curve has genus 1, i.e., it is an elliptic curve ${\cal E}_0$.
It is parametrized by points $z$ of
one-dimensional complex torus with periods $1$ and $\tau$.

Fixing a suitable normalization (for example, putting second components
of all the vectors equal to 1), we may consider components of the
vectors $U(z),\,V(z),\,X(z),\,Y(z)$ as meromorphic functions on
${\cal E}_0$ having at most 2 simple poles. With this normalization,
the right hand side of (\ref{5.1a}) must be multiplied by a scalar
meromorphic function $h(z)$. It follows \cite{Krichvac}
from the Yang-Baxter equation that
\beq
Y(z)=X(z+\frac{\eta}{2}),\;\;\;V(z)=U(z-\frac{\eta '}{2}),
\label{5.1d}
\eeq
where $\eta$ and $\eta '$ are some constants. Therefore, we can write
the basic relation (\ref{5.1a}) in the form \cite{Krichvac}
\beq
{\cal L}(X(z)\otimes U(z))\!=\!h(z)Y(z)\otimes V(z)\!=\!
h(z)X(z+\frac{\eta}{2})\otimes U(z-\frac{\eta '}{2}).
\label{5.1c}
\eeq

Let $D_X$ (resp., $D_U$) be the divisor of poles of the meromorphic vector
$X(z)$ (resp., $U(z)$). Let ${\cal M}(D)$ be the space of functions
associated to an effective divisor $D$, i.e., functions having poles
at points of $D$
of order not higher than the multiplicity of the corresponding point
in $D$. For divisors of degree 2 on elliptic curves
this space is two-dimensional
in general position, so $\dim {\cal M}(D_X)=
\dim {\cal M}(D_U)=2$, and components of the vectors $X$ and $U$ form
bases in these spaces. Further, the functions $X_i (z)U_{\alpha}(z),\,\,
i,\alpha =1,2 $, form a basis in the space ${\cal M}(D_X +D_U)$. According
to (\ref{5.1c}), the functions $h(z)X_i (z+\frac{\eta}{2})
U_{\alpha}(z-\frac{\eta '}{2})$ form another basis in this space and
the matrix ${\cal L}$ connects the two bases.
The divisors of poles of the left
and right hand sides of (\ref{5.1c}) must be equivalent, i.e., must
be equal modulo periods of the lattice. Since under the shift by
$\eta '/2$ the divisor of poles of the function having 2 poles is
shifted by $\eta '$, this means that
\beq
\eta '-\eta =M+N\tau\,, \;\;\;\;\;M,N\in {\bf Z}.
\label{5.8}
\eeq
The vectors $X(z)$, $U(z)$ are double-periodic, hence there are
4 different cases:
\beq
\eta ' = \eta +2\omega _a\,,
\label{5.9}
\eeq
where $\omega _0 =0$, $\omega _1=1/2$, $\omega _2=(\tau +1)/2$,
$\omega _3=\tau /2$ (cf. (\ref{g201})).

Baxter's parametrization of the $L$-operator follows
from eq. (\ref{5.1c}). Indeed, the equivalence class of the
pole divisor of $X(z)$ may differ from that of $U(z)$ by
only a shift on ${\cal E}_0$. By means of a "gauge" transformation
one may represent ${\cal L}$ in the form (\ref{4}).
 The value of this shift is then identified with the spectral parameter
of the $L$-operator. In this parametrization, it is natural to
write (\ref{5.1c})
explicitly in terms of $\theta$-functions.

To do this, it is convenient to use another normalization, specifically,
the one in which the vectors are entire functions in $z$ (in this case
they are sections of certain line bundles on ${\cal E}_0$).

Let us introduce the vector
\beq
\Theta(z)= \left ( \begin{array}{c}
\theta_{4}(z|\displaystyle{\frac{\tau}{2}})\\
\phantom{a}\\
\theta_{3}(z|\displaystyle{\frac{\tau}{2}}) \end{array} \right ).
\label{5.5}
\eeq
Its components form a basis in the space of $\theta$-functions
$\theta (z)$ of
second order with monodromy propertties $\theta (z+1)=\theta (z)$,
$\theta (z+\tau)=\exp (-2\pi i \tau -4\pi iz)\theta (z)$.
They have 2 zeros in the fundamental domain of the
lattice with periods $1\,,\tau$.
Then, putting
\beq
X(z)=\Theta (z)\,,
\label{5.10a}
\eeq
\beq
U(z)=U^{\pm }(z)=\Theta \left (z\pm \frac{1}{2}
(u+\frac{\eta}{2})\right ) \label{5.10b} \eeq
(for each choice of the sign),
we may rewrite (\ref{5.1a}) in the form \cite{Baxter},
\cite{ft}:
\beq
L^{(a)}(u)\Theta(z)\otimes
\Theta(z\pm \frac{1}{2}(u+\frac{\eta}{2}))=
2g_{a}^{\pm}\frac{\theta_1(u+\frac{\eta}{2}|\tau)}
{\theta_1(\eta |\tau)}\!
\Theta (z+\frac{\eta}{2})\otimes  \Theta
(z\pm \!\frac{1}{2}(u-\frac{\eta }{2})\pm \omega _a),
\label{5.6}
\eeq
where
\beq
g_{0}^{\pm}=g_{1}^{\pm}=1,\,\,\,\,\,
-ig_{2}^{\pm}=g_{3}^{\pm}=-\exp (\pm 2\pi iz+\pi i(u+
\frac{\tau -\eta }{2})),
\label{5.6a} \eeq
\beq
L^{(a)}(u)=\sum _{b=0}^{3}\frac{ \theta _{b+1}(u|\tau)}
{\theta _{b+1}(\frac{\eta}{2}|\tau)} \sigma _{b}\otimes
(\sigma _{a}\sigma _b )
\label{5.7}
\eeq
(see (\ref{4}) and (\ref{6})). We note that
$L^{(a)}(u)=\sigma _a L(u)$ ($L(u)$ is given by (\ref{4}) with
$S_a =\sigma _a$ and the matrix product is performed in the
auxiliary space) satisfies the "$RLL=LLR$" relation (\ref{5}) with
{\it the same} $R$-matrix (\ref{6}) for each $a=0,\ldots,3$.
The scalar factor in the right hand side of
(\ref{5.6}) is determined from the condition that $L(\frac{\eta}{2})$
is proportional to the permutation operator in ${\bf C}^2 \otimes
{\bf C}^2$. One may verify (\ref{5.6}) directly using identities for
$\theta$-functions (see the Appendix).

{\bf Remark}. Given an elliptic curve, we can always choose the
vector $X(z)$ to be an even function, $X(-z)=X(z)$. We introduce the even
function $X(z)$ (\ref{5.5}) from the very beginning. Then the
equality corresponding to the minus sign in (\ref{5.6}) follows from
the similar equality with the plus
(it is enough to change $z\rightarrow -z$).
However, in Baxter's approach it is useful to
deal with both equalities.

Let us turn to the case of arbitrary spin. Consider an
$L$-operator of the form (\ref{4}):
\beq
L(u)=\left ( \begin{array}{cc}
W_0 (u)S_0 +W_3 (u)S_3 & W_1 (u)S_1-iW_2 (u)S_2 \\
W_1 (u)S_1 +iW_2 (u)S_2 & W_0 (u)S_0-W_3 (u)S_3 \end{array} \right ),
\label{5.10} \eeq
where
\beq
W_a (u)=\frac {\theta _{a+1}(u|\tau)}{\theta _{a+1}(\frac{\eta}{2}|\tau)}\,,
\label{5.11} \eeq
and $S_a$ are generators of some algebra (at this stage the commutation
relations (\ref{1}), (\ref{2}) are not imposed). We are going to
obtain explicit formulas for representations of this algebra from a simple
generalization of (\ref{5.1c}).

Before presenting the main result of this section we need
some more preliminaries.

Since we are interested in the general form of
representations in terms of difference
operators,
in what follows we take for the quantum space of the $L$-operator
(\ref{5.10}) the space of {\it meromorphic functions of one complex
variable} $z$. The generators $S_a$ act on elements of this space
(i.e. functions $X(z)$):
\beq
S_a\,:\, X(z)\longmapsto (S_a X)(z) \,.
\label{5.12} \eeq

Consider the following generalization of (\ref{5.1c}) and (\ref{5.6}):
\beq
U^{\pm}(z)^{T}(\sigma _a L(u))X(z)=g_{a}(u)U^{\pm}
(z\mp \ell \eta \mp \omega _a)^{T}
X(z\pm \frac{\eta}{2}),\,\,\,\,\,a=0,\ldots,3.
\label{5.13} \eeq
Here $z\in {\cal E}_0$, $U^{T}=(U_1 , U_2 )$,
$\ell$ is a parameter, $g_{a}(u)$ are scalar functions independent of
$z$, and \beq U^{\pm}(z)=\Theta \left (z\pm \frac{u+\ell \eta}{2}\right )
\label{5.14} \eeq
(for $\ell =1/2$ it coincides with (\ref{5.10b})). As before, $L(u)$ acts
on $U^{\pm}$ as 2$\times$2-matrix, while each matrix element of $L(u)$
acts on $X(z)$ according to (\ref{5.12}). We have written (\ref{5.13})
in terms of the {\it covector} $U^{T}$ for the following reason:
since the operator (\ref{5.12}) acts on a function {\it from the left},
this is equivalent to the {\it right} action of the
corresponding matrix (representing this
operator in a fixed basis) on the covector formed by components of the
function with respect to the basis. As it is clear from what follows,
at $\ell =1/2$ (\ref{5.13}) coincides with the {\it conjugated} equality
(\ref{5.6}).

It should be noted that at present time we can not suggest any explicit
description of the vacuum curve of the $L$-operator (\ref{5.10}).
Moreover, we do not know any direct argument
establishing the equivalence between (\ref{5.13}) and the "intertwining"
relation (\ref{5}) for $L(u)$ (taken together with the Yang-Baxter
equation for $R(u)$). Theorem 6.1 (see below) states that Sklyanin's
commutation relations (\ref{1}, \ref{2}) for $S_a$ follow from
(\ref{5.13}), hence $L(u)$ should satisfy (\ref{5}). Let us stress
once again that our arguments are in a sense inverse to original
Sklyanin's approach (see also \cite{Tak}, \cite{Tak1}, where some
formulas for vacuum vectors in the higher spin $XYZ$ model were obtained).
Our starting point is the relation (\ref{5.13}), where no conditions
on $S_a$ are implied. It turns out that the Sklyanin algebra for
$S_a$ (together with its functional realization)
follows from (\ref{5.13}).

The main result of this section is the following
\begin{th}
Let $L(u)$ be given by (\ref{5.10}), where $S_a$ are some operators
in the space of meromorphic functions $X(z)$ of one complex variable
$z$. Suppose the relation (\ref{5.13}) holds for $a=0$, i.e.,
\beq
U^{\pm}(z)^{T}L(u)X(z)=g_{0}(u)U^{\pm}
(z\mp \ell \eta )^{T}
X(z\pm \frac{\eta}{2}),
\label{5.15} \eeq
where $U^{\pm }(z)$ is defined in (\ref{5.14}), $l$ is a parameter
and $g_0 (u)$ is a scalar function independent of $z$. Then $S_a$
are difference operators of the following form:
\beq
(S_a X)(z)=\lambda
\frac{ (i)^{\delta _{a,2}} \theta _{a+1}(\frac{\eta}{2}|\tau)}
{\theta _{1}(2z|\tau)}\left ( \theta _{a+1}(2z-\ell \eta |\tau)
X(z+\frac{\eta}{2})-\theta _{a+1}(-2z-\ell \eta |\tau)
X(z-\frac{\eta}{2})\right ),
\label{5.16} \eeq
where $\lambda$ is an arbitrary constant. Conversely, if $S_a$ are
defined by (\ref{5.16}), then (\ref{5.15}) holds, and
\beq
g_0 (u)=2\lambda \theta _{1}(u+\ell \eta |\tau)\,.
\label{5.17} \eeq
\end{th}

{\bf Remark}. a) Using transformation properties of the vector
$\Theta (z)$ under shifts by half-periods $\omega _a$ it is readily
seen that the relations (\ref{5.13}) for $a\ne 0$ follow from
(\ref{5.15}); b) For even functions $X(z)$ the two relations (\ref{5.15})
are equivalent.

{\it Proof.} The relations (\ref{5.15}) form a system of 4 linear
equations for 4 functions $(S_a X)(z)$ entering the left hand side.
More explicitly, we have
\beq
U_{1}^{\pm}(z)(L_{11}(u)X)(z)+U_{2}^{\pm}(z)(L_{21}(u)X)(z)=
g_0 (u)U_{1}^{\pm}(z\mp \ell \eta)X(z\pm \frac{\eta}{2})\,,
\eeq
\beq
U_{1}^{\pm}(z)(L_{12}(u)X)(z)+U_{2}^{\pm}(z)(L_{22}(u)X)(z)=
g_0 (u)U_{2}^{\pm}(z\mp \ell \eta)X(z\pm \frac{\eta}{2})\,,
\eeq
where $(L_{\alpha \beta}(u)X)(z)$ are expressed through $(S_a X)(z)$
according to (\ref{5.10}). Fix $g_0 (u)$ to be given by (\ref{5.17}).
Solving this system, we get (\ref{5.16}) (all necessary identities for
$\theta$-functions are presented in the Appendix to this section).
Therefore, (\ref{5.16}) is equivalent to (\ref{5.15}, \ref{5.17}).
\square

\begin{cor}
Suppose for some $L(u)$ of the form (\ref{5.10}) the equation
(\ref{5.15}) holds. Then $L(u)$ satisfies the "intertwining" relation
(\ref{5}) with the $R$-matrix (\ref{6}).
\end{cor}
The proof follows from the identification of (\ref{5.16}) with
formulas (\ref{10}) for representations of the Sklyanin algebra by
putting $2z \equiv x$, $X(x/2)\equiv f(x)$. The constant $\lambda$ is
not essential since the commutations relations
(\ref{1}, \ref{2}) are homogeneous.

{\bf Remark}. From the technical point of view, the derivation of
formulas (\ref{10}) by solving the system (\ref{5.15}) is much
simpler than the direct verification of the commutation relations.
The amount of computations in the former case is comparable with
that in the latter one if we identify the coefficients
only in front of $f(x\pm 2\eta)$.

Let us consider now the equality (\ref{5.13}) for $a=1,2,3$.

\begin{lem}
Let us define the transformation of the generators $S_b \longmapsto
{\cal Y}_a (S_b )$, $a,b =0,\ldots, 3$, by the relation
\beq
\sigma _a L(u+\omega _a)=h_a (u)
\sum _{b=0}^{3}\frac {\theta _{b+1}(u|\tau)}
{\theta _{b+1}(\frac{\eta}{2}|\tau)} {\cal Y}_a (S_b )\otimes
\sigma _b \,,
\label{5.18} \eeq
where $h_0 (u)=h_1 (u)=1$, $h_2 (u)=-ih_3 (u)=\exp (-\frac{i\pi \tau}
{4}-i\pi u)$. Then ${\cal Y}_a $, $a=0,\ldots, 3$, is an automorphism
of the algebra generated by $S_b$.
\end{lem}
The proof follows from the fact that $\sigma _a L(u+\omega _a)$,
$a=0,\ldots, 3$ satisfies (\ref{5}) with $R$-matrix (\ref{6}).
(This is because the matrices $\sigma _a$ are $c$-number solutions
of (\ref{5}).) From (\ref{5.18}) we find the explicit form of
${\cal Y}_a$ (here $\theta _a(\frac{\eta}{2})\equiv \theta _a
(\frac{\eta}{2}|\tau)$):
\beq
{\cal Y}_1 :(S_0,S_1,S_2,S_3)\longmapsto
\left ( -\frac{\theta _1(\frac{\eta}{2})}{\theta _2(\frac{\eta}{2})}S_1,\,
\frac{ \theta _2(\frac{\eta}{2})}{\theta _1(\frac{\eta}{2})}S_0, \,
-\frac{i \theta _3(\frac{\eta}{2})}{\theta _4(\frac{\eta}{2})}S_3, \,
\frac{i\theta _4(\frac{\eta}{2})}{\theta _3(\frac{\eta}{2})}S_2 \right ),
\label{Y1} \eeq
\beq
{\cal Y}_2 :(S_0,S_1,S_2,S_3)\longmapsto
\left ( \frac{\theta _1(\frac{\eta}{2})}{\theta _3(\frac{\eta}{2})}S_2,\,
\frac{i \theta _2(\frac{\eta}{2})}{\theta _4(\frac{\eta}{2})}S_3, \,
\frac{ \theta _3(\frac{\eta}{2})}{\theta _1(\frac{\eta}{2})}S_0, \,
-\frac{\theta _4(\frac{\eta}{2})}{\theta _2(\frac{\eta}{2})}S_1 \right ),
\label{Y2} \eeq
\beq
{\cal Y}_3 :(S_0,S_1,S_2,S_3)\longmapsto
\left ( \frac{\theta _1(\frac{\eta}{2})}{\theta _4(\frac{\eta}{2})}S_3,\,
-\frac{ \theta _2(\frac{\eta}{2})}{\theta _3(\frac{\eta}{2})}S_2, \,
\frac{ \theta _3(\frac{\eta}{2})}{\theta _2(\frac{\eta}{2})}S_1, \,
\frac{\theta _4(\frac{\eta}{2})}{\theta _1(\frac{\eta}{2})}S_0 \right ),
\label{Y3} \eeq
and ${\cal Y}_0$ is the identity transformation. These automorphisms were
considered by Sklyanin in \cite{skl2}.

Shifting $u\rightarrow u+\omega _a$ in (\ref{5.13}) and
applying these automorphisms to (\ref{5.16}), we get for each
$b=0,\ldots,3$ the following representations:
\beq
\!\!(S_a X)(z)\!=\!
\frac{(i)^{\delta _{a,2}} \theta _{a+1}(\frac{\eta}{2})}
{\theta _{1}(2z)}\!
\left (\! \theta _{a+1}(2z\!-\!\ell \eta \!-\!\omega _b)
X(z+\frac{\eta}{2})\!-\!\theta _{a+1}(-2z\!-\!\ell \eta \!-\!\omega _b)
X(z-\frac{\eta}{2})\!\right ).
\label{5.19} \eeq
We remark that (\ref{5.19}) can be obtained from (\ref{5.16}) by the
formal change $\ell \rightarrow \ell +\omega _b/ \eta$ provided $\ell$
is considered to be an arbitrary complex parameter. However, we will see
soon that the operators (\ref{5.19})
restricted to invariant subspaces
yield non-equivalent finite-dimensional
representations of the Sklyanin algebra.

Suppose $\ell \in \frac{1}{2}{\bf Z}_{+}$; then one may identify $X(z)$
with a section of some linear bundle on the initial elliptic
curve ${\cal E}_0$. Repeating the arguments presented after eq.
(\ref{5.1c}), we conclude that degree of this bundle equals $4\ell$.
{}From the Riemann-Roch theorem for elliptic curves it follows that
in general position the space of holomorphic sections of this bundle
is $4\ell$-dimensional. It is convenient to identify these sections with
$\theta$-functions of order $4\ell$. Consider the space ${\cal
T}^{+}_{4\ell}$ of {\it even} $\theta$-functions of order $4\ell$, i.e.,
the space of entire functions $F(z)$, $z\in {\bf C}$, such that
$F(-z)=F(z)$ and \beq \begin{array}{l} F(z+1)=F(z)\,,\\ F(z+\tau)=\exp
(-4\ell \pi i \tau -8\ell \pi i z)F(z)\,.\\
\end{array} \label{5.20} \eeq
In complete analogy with the paper \cite{skl2}, one can see that the
space ${\cal T}_{4\ell}^{+}$ is invariant under action of the operators
(\ref{5.19}) for $b=0$ and $b=1$, while for $b=2$ and $b=3$ the
invariant space is $\exp (-\pi i z^{2}/\eta ){\cal T}_{4\ell}^{+}$.  It
is known that $\dim {\cal T}_{4\ell}^{+}=2\ell +1$ provided $\ell \in
\frac{1}{2}{\bf Z}_{+}$. Restricting the difference operators
(\ref{5.19}) to these invariant subspaces, we get 4 series of
finite-dimensional representations.

Generally speaking, these representations are mutually non-equivalent.
This follows from analysis of values of the central elements. The
Sklyanin algebra has two independent central elements:
\beq
{\bf K}_0 =\sum _{a=0}^{3}S_{a}^2\,,\;\;\;\;
	{\bf K}_2 =\sum _{i=1}^{3}J_i S_{i}^2
	\label{center} \eeq
(the constants $J_i$ are defined by (\ref{3}) and (\ref{13})). Their
values for the representations (\ref{5.19}) at $b=0,\ldots,3$ are:
\beq
{\bf K}_0 =4\theta _{1}^{2}\left ((\ell +\frac{1}{2})\eta +\omega _b
|\tau \right ),
\label{center0} \eeq
\beq
{\bf K}_2 =4\theta _1 \left ((\ell +1)\eta +\omega _b |\tau\right )
\theta _1 \left (\ell \eta +\omega _b |\tau \right ).
\label{center2} \eeq
The arguments given above lead to the following statement.
\begin{th}
The Sklyanin algebra (\ref{1}, \ref{2}) has 4 different series of
finite-dimensional representations indexed by
$b=0,\,1,\,2,\,3$. Representations of each series
are indexed by the discrete parameter $\ell \in \frac{1}{2}{\bf Z}_{+}$
("spin"). They are obtained by restriction of the operators (\ref{5.19})
to the invariant $(2\ell +1)$-dimensional functional subspaces
${\cal T}_{4\ell}^{+}$ for $b=0,\,1$ and
$\exp (-\pi i z^{2}/\eta){\cal T}_{4\ell}^{+}$ for $b=2,\,3$. For general
values of parameters these representations are mutually non-equivalent.
\end{th} \square

Let us identify these representations with those obtained by
Sklyanin in \cite{skl2}. For $b=0$ and $b=3$ we get series a)
and c) respectively. These representations are self-adjoint with
respect to the real form of the algebra studied in \cite{skl2}.
The other two series (corresponding to $b=1$ and $b=2$) in
general are not
self-adjoint. Consider first the case $b=1$.
For rational values of $\eta$, $\eta =p/q$, and special values of $\ell$,
$\ell= (q-1)/2\!$ mod $\!q$,
these representations are self-adjoint and are equivalent to
some subset of representations of series b) \footnote{The whole family
of representations of series b) found in \cite{skl2} has 3
continuous parameters. These representations
are self-adjoint and exist only if
$\eta =p/q$. In this case all of them have dimension $q$. They
are obtained by restriction of the operator
(\ref{5.19}) to a finite discrete uniform lattice.}.
To the best of our knowledge, the series corresponding to $b=2$ was
never mentioned in the literature (though, in a sense, it is implicitly
contained in Sklyanin's paper). Another outcome of our approach is
the natural correspondence between the different series
of representations and points of order 2
on the elliptic curve.

It is natural to surmise that representations of the last two series
become self-adjoint with respect to other real forms of the
algebra. A real form is defined by an anti-involution ($*$-operation)
on the algebra. It should be noted that classification of
non-equivalent real forms of the Sklyanin algebra and its generalizations
is an interesting open problem.

Concluding this section, we would like to remark that the variable
$z$ in (\ref{5.13}) and (\ref{5.19}) may be identified with the
statistical variable ("height") in IRF-type models \cite{ABF}
(after a suitable discretization of the former). This is readily seen
from the well known vertex-IRF correspondence recalling that
the vertex-IRF transformation is performed by means of
the vacuum vectors (for the explicit form of this transformation in
the case of higher spin models see \cite{Tak1}).

Finally, it seems to be instructive to carry out a detailed analysis
of the trigonometric and rational limits
of the constructions presented in this
section. Some particular related problems have been already discussed
in the literature.
In the recent paper \cite{YuBa}, vacuum vectors for the higher spin
$XXZ$-type quantum spin chains are constructed. Vacuum curves of
trigonometric $L$-operators have been described in \cite{Kor}.
In the simplest case they are
collections of rational curves intersecting at 2 points.
Trigonometric degenerations of the Sklyanin algebra that are in a
sense "intermediate" between the initial algebra and the standard
quantum deformation of $gl_2$ are studied in \cite{GZ}.

\subsubsection*{Appendix to Section 6}

We use the following definition of the $\theta$-functions:
\beq
\theta _1(z|\tau)=\sum _{k\in {\bf Z}} \exp \left (
\pi i \tau (k+\frac{1}{2})^2 +2\pi i
(z+\frac{1}{2})(k+\frac{1}{2})\right ),
\label{def1} \eeq
\beq
\theta _2(z|\tau)=\sum _{k\in {\bf Z}} \exp \left (
\pi i \tau (k+\frac{1}{2})^2 +2\pi i
z(k+\frac{1}{2})\right ),
\label{def2} \eeq
\beq
\theta _3(z|\tau)=\sum _{k\in {\bf Z}} \exp \left (
\pi i \tau k^2 +2\pi i
zk \right ),
\label{def3} \eeq
\beq
\theta _4(z|\tau)=\sum _{k\in {\bf Z}} \exp \left (
\pi i \tau k^2 +2\pi i
(z+\frac{1}{2})k\right ).
\label{def4} \eeq
For reader's convenience we recall here the definition of the
$\sigma$-function used in Sects.~2~--~4:
\beq
\sigma (z|\omega ,\omega ')=\frac{2\omega }{\theta_{1}'(0)}
\exp \left (\frac{ \zeta (\omega )z^2}{2\omega }\right )
\theta _1 (\frac{z}{2\omega }\,|\,\frac{\omega '}{\omega})\,.
\label{sigma} \eeq

Here is the list of identities used in the computations.

The first group of identities (addition theorems):
\beq
\theta _4 (x|\tau)\theta _3(y|
\tau)=\theta_4 (x+y|2\tau)
\theta _4(x-y|2\tau)+\theta _1(x+y|2\tau)\theta _1(x-y|2\tau),
\label{theta1} \eeq
\beq
\theta _4 (x|\tau)\theta _4(y|\tau)=\theta_3 (x+y|2\tau)
\theta _3(x-y|2\tau)-\theta _2(x+y|2\tau)\theta _2(x-y|2\tau),
\label{theta2} \eeq
\beq
\theta _3 (x|\tau)\theta _3(y|\tau)=\theta_3 (x+y|2\tau)
\theta _3(x-y|2\tau)+\theta _2(x+y|2\tau)\theta _2(x-y|2\tau),
\label{theta3} \eeq
\beq
\theta _2 (x|\tau)\theta _2(y|\tau)=\theta_3 (x+y|2\tau)
\theta _2(x-y|2\tau)+\theta _2(x+y|2\tau)\theta _3(x-y|2\tau),
\label{theta4} \eeq
\beq
\theta _1 (x|\tau)\theta _1(y|\tau)=\theta_3 (x+y|2\tau)
\theta _2(x-y|2\tau)-\theta _2(x+y|2\tau)\theta _3(x-y|2\tau),
\label{theta5} \eeq

Here are simple consequences of them convenient in the computations:
\beq
\theta _4 (x|\tau)\theta _3(y|\tau)+\theta _4 (y|\tau)\theta _3(x|\tau)=
2\theta _4 (x+y|2\tau)\theta_4 (x-y|2\tau),   \label{theta6} \eeq
\beq
\theta _4 (x|\tau)\theta _3(y|\tau)-\theta _4 (y|\tau)\theta _3(x|\tau)=
2\theta _1 (x+y|2\tau)\theta_1 (x-y|2\tau),   \label{theta7} \eeq
\beq
\theta _3 (x|\tau)\theta _3(y|\tau)+\theta _4 (y|\tau)\theta _4(x|\tau)=
2\theta _3 (x+y|2\tau)\theta_3 (x-y|2\tau),   \label{theta8} \eeq
\beq
\theta _3 (x|\tau)\theta _3(y|\tau)-\theta _4 (y|\tau)\theta _4(x|\tau)=
2\theta _2 (x+y|2\tau)\theta_2 (x-y|2\tau).   \label{theta9} \eeq

The second group of identities:
\beq
2\theta _1 (x|2\tau)\theta _4 (y|2\tau)=
\theta _1 (\frac{x+y}{2}|\tau)\theta _2 (\frac{x-y}{2}|\tau)+
\theta _2 (\frac{x+y}{2}|\tau)\theta _1 (\frac{x-y}{2}|\tau),
\label{theta10} \eeq
\beq
2\theta _3 (x|2\tau)\theta _2 (y|2\tau)=
\theta _1 (\frac{x+y}{2}|\tau)\theta _1 (\frac{x-y}{2}|\tau)+
\theta _2 (\frac{x+y}{2}|\tau)\theta _2 (\frac{x-y}{2}|\tau),
\label{theta11} \eeq
\beq
2\theta _3 (x|2\tau)\theta _3 (y|2\tau)=
\theta _3 (\frac{x+y}{2}|\tau)\theta _3 (\frac{x-y}{2}|\tau)+
\theta _4 (\frac{x+y}{2}|\tau)\theta _4 (\frac{x-y}{2}|\tau),
\label{theta12} \eeq
\beq
2\theta _2 (x|2\tau)\theta _2 (y|2\tau)=
\theta _3 (\frac{x+y}{2}|\tau)\theta _3 (\frac{x-y}{2}|\tau)-
\theta _4 (\frac{x+y}{2}|\tau)\theta _4 (\frac{x-y}{2}|\tau).
\label{theta13} \eeq

Particular cases of them:
\beq
2\theta _1 (z|\tau)\theta _4 (z|\tau)=\theta _2 (0|\frac{\tau}{2})
\theta _1 (z|\frac{\tau}{2}), \label{theta14} \eeq
\beq
2\theta _2 (z|\tau)\theta _3 (z|\tau)=\theta _2 (0|\frac{\tau}{2})
\theta _2 (z|\frac{\tau}{2}). \label{theta15} \eeq

Two more identities:
\beq
\theta _1 (z|\frac{\tau}{2})\theta _2 (z|\frac{\tau}{2})=
\theta _4 (0|\tau)\theta _1 (2z|\tau), \label{theta16} \eeq
\beq
\theta _4 (z|\frac{\tau}{2})\theta _3 (z|\frac{\tau}{2})=
\theta _4 (0|\tau)\theta _4 (2z|\tau). \label{theta17} \eeq

\bigskip
\section{Concluding remarks}

This work elaborates upon the following three subjects:

I) Dynamics of poles for elliptic solutions to the 2D non-abelian
Toda chain;

II) Difference analogs of Lame operators;

III) Representations of the Sklyanin algebra in terms of difference
operators.

Let us outline the results:

- The poles move according to equations of motion for spin generalizations
of the Ruijsenaars-Schneider model; the action-angle variables for the
latter are constructed in terms of some algebraic-geometrical data;

- One of the generators of the Sklyanin algebra, represented as
a difference operator with elliptic coefficients, has the
"finite-gap" property that is a motivation for the analogy with
Lame operators;

- Starting from the notion of vacuum vectors of an $L$-operator,
a general simple scheme for constructing functional realizations of the
Sklyanin algebra is suggested.

Here we would like to explain why the three themes are to be
intimately connected.

To each problem I) - III), a distinguished class of algebraic curves
has been associated. In case I), these are spectral curves $\G$ for
the $L$-operators of the Ruijsenaars-Schneider-type models;
in II), we deal with the spectral curve $\G '$ for the difference
Lame operator $S_0$ (a generator of the Sklyanin algebra);
in III), the representations are defined on sections of certain
line bundles on a vacuum curve ${\cal E}$ of the elliptic higher
spin $L$-operator (\ref{4}) (though it is implicit in Section 6).
It has been shown that $\G$ and $\G '$ are ramified coverings of
the initial elliptic curve. The characteristic property (\ref{5.13})
of the vacuum vectors suggests that the same should be true for
${\cal E}$, i.e., ${\cal E}$ is a ramified covering of the initial
elliptic curve ${\cal E}_0$ (the vacuum curve of the spin-1/2
$L$-operator).

The connection between I) and II) is similar to the relation between
elliptic solutions of KP and KdV equations. Specifically, the
elliptic solutions of the abelian 2D Toda chain, which are
{\it stationary with respect to the time flow} $t_{+}+t_{-}$
correspond to {\it isospectral deformations} of the difference
Lame operator $S_0$ (considered as a Lax operator for 1D Toda
chain). In other words, the hyperelliptic
curves $\G '$ form a specific subclass of the curves $\G$.
A similar reduction in the non-abelian case yields spin generalizations
of difference Lame operators. Their properties and a possible relation
to the Sklyanin-type quadratic algebras are to be figured out.

Apart from the apparent result that the construction of Sect. 6
provides a natural source of difference Lame-like operators, we
expect a more deep connection between II) and III). Specifically,
the spectral curves $\G '$ are expected to be very close to the
vacuum curves ${\cal E}$. Conjecturally, they may even coincide, at least
in some particular cases. At the moment we can not present any more
arguments and leave this as a further problem.

At last, we would like to note an intriguing similarity between the
basic ansatz (\ref{1.7}) for a double-Bloch solution of the
generating linear problem and the functional
Bethe ansatz \cite{skl3}. Indeed, in the latter case wave functions
are sought in the form of an "elliptic polynomial" $\prod \sigma (z-z_j )$,
where the roots $z_j$ are subject to Bethe equations. Similarly,
in the former case we deal with {\it a ratio} of two "elliptic
polynomials" (cf. (\ref{g21})).
However, this function is parametrized by residues at
the poles rather than zeros of the numerator. This may indicate a
non-trivial interplay \footnote{The recently observed formal
resemblance \cite{NRK} between Bethe equations and
equations of motion for discrete time Calogero-Moser-like
systems may be a particular aspect of this relation.}
between Calogero-Moser-type models (and more general
Hithin's systems) and quantum integrable models solved by means
of Bethe ansatz.

\bigskip

\section*{Acknowledgements}
We are grateful to A.Gorsky and T.Takebe
for discussions. The work of I.K. was
supported by grants ISF MD-8000 and
grant 93-011-16087 of Russian Foundation of Fundamental Research. I.K.
is very grateful to Technion University at Haifa, Rome1 University,
Scoula Normale at Pisa and Berlin Technishe University for their
hospitality during periods when this work was done.
The work of A.Z. was supported in part by
grant 93-02-14365 of Russian Foundation of Fundamental Research, by
grant ISF MGK300 and by ISTC grant 015.

\end{document}